\documentclass[12pt,onecolumn,draftcls]{IEEEtran}
\pdfoutput=1
\usepackage[pdftex]{graphicx}

\usepackage{epsfig}

\usepackage{subfigure}

\usepackage{amssymb}

\usepackage{amsbsy}

\usepackage{amsmath}

\usepackage{cite}

\usepackage{float}

\usepackage{bbm}

\usepackage{url}

\usepackage{multirow}

\newtheorem{theorem}{Theorem}

\newtheorem{corollary}{Corollary}

\begin{document}

\title{Asynchronous Transmission over Single-User State-Dependent Channels}

\author{Michal Yemini$^{*}$, Anelia Somekh-Baruch$^{*}$ and Amir Leshem$^{*}$

\thanks{The paper ``On channels with asynchronous side information" was  split
into two separate papers: the enclosed  paper which considers only point-to-point
channels and an additional paper named ``On the multiple access channel with asynchronous cognition" which discusses the multiuser setups.}
\thanks{This research was partially supported by Israel Science Foundation (ISF) grant  2013/919.}
\thanks{The results of this paper were partially presented in \cite{Eilat_us}.}
\thanks{$*$ Faculty of Engineering, Bar-Ilan University, Ramat-Gan, 52900, Israel. Email:
michal.yemini.biu@gmail.com, anelia.somekhbaruch@gmail.com, leshem.amir2@gmail.com.}
}
\date{}

\maketitle

\begin{abstract}
Several channels with asynchronous side information are introduced.
We first consider single-user state-dependent channels with asynchronous side information at the transmitter. It is assumed that the state information sequence is a possibly delayed version of the state sequence, and that the encoder and the decoder are aware of the fact that the state information might be delayed. It is additionally assumed that an upper bound on the delay is known to both encoder and decoder, but other than that, they are ignorant of the actual delay. We consider both the causal and the noncausal cases and present achievable rates for these channels, and the corresponding coding schemes.  We find the capacity of the asynchronous Gel'fand-Pinsker channel with feedback. Finally, we consider a memoryless state dependent channel with asynchronous side information at both the transmitter and receiver, and establish a single-letter expression for its capacity.
\end{abstract}

\begin{IEEEkeywords}
Asynchronism, binning, causal side information, channel capacity, channel coding, cognitive radio, Gel'fand-Pinsker channel, non-causal side information, strategy letters.
\end{IEEEkeywords}

\section{Introduction}
State dependent channels with side information known at the encoder were first introduced by Shannon.
In \cite{shannon}, Shannon established a single-letter expression for the capacity of state dependent channels with side information known causally at the encoder and unknown to the decoder. Subsequently, Kusnetsov and Tsybakov \cite{Kusnetsov} introduced channels with i.i.d. side information which is known non-causally at the encoder and Gel'fand and Pinsker derived the formula for the capacity of these channels using random binning encoding methods \cite{GP}.

The introduction of state dependent channels with side information at the transmitter was originally aimed at analyzing coding techniques for computer memory with defect whose locations are known to the encoder only \cite{Heegard}. With the development of communication systems and the Internet, other relevant applications have emerged. Amongst them are cognitive radio \cite{Devroye,GoldsmithJafar2009}, watermarking \cite{Moulin}, multiple-input multiple-output broadcast channels \cite{Caire}, multiple-access channels with channel side information \cite{Anelia2008}, etc.
The common underlying assumption in the analysis of these channels, is that the side information signal is synchronized with the signal produced by the encoder. However, in practical situations this assumption does not necessarily hold and the side information signal may be a delayed version of the channel states sequence. When the assumption of the synchronization does not hold, the known results of the aforementioned channels are not necessarily valid, new models that encompass the unknown delay of the state sequence at the transmitter need to be addressed.

Other models that may suffer from asynchronism are multi-user channels, in which users are assumed to be synchronized with  one another. The discrete memoryless multiple access channel (MAC) with independent sources was the first channel from this family that was considered in an asynchronous setup \cite{CoverMcEliece1981,HuiHumblet1985,Verdu1989}. It was shown by Cover \textsl{et. al.} \cite{CoverMcEliece1981} that if the delay is finite or grows sufficiently slowly relatively to the block length, then the asynchronism does not change the capacity region. However, Hui and Humblet \cite{HuiHumblet1985} showed that the capacity region may be reduced if the delay is of the same order of the block length, since time sharing cannot be used.

In this paper, we address the question of whether an asynchronous side information is useful when the delay is bounded. By lower bounding the achievable rates using time sharing between all possible delays, we prove that the asynchronous side information can still be of value in the asynchronous Gel'fand-Pinsker channel \cite{Eilat_us}. We improve the lower bound for the asynchronous Gel'fand-Pinsker channel  by studying two of its counterparts: the multicast channel \cite{Khisti}, and the compound channel \cite{Piantanida,Nair}, and by taking into account the specific characteristics of our setup. In addition, we observe that if feedback is present, the capacity of the asynchronous Gel'fand-Pinsker channel is equal to the capacity of the synchronous Gel'fand-Pinsker channel.
We additionally consider state dependent channels with state information available asynchronously and causally at the transmitter. Contrary to the non-causal and asynchronous state information, in the causal setup there are cases in which the side information does not improve the reliably transmitted rates. We distinguish between two cases of possible delay values:  If the maximal delay is positive, i.e., the encoder may observe at each time instant a past actual state, then the side information can be ignored without loss of optimality. Otherwise, a scheme which is based on the limited lookahead scheme of \cite{WeissmanElGamal} is presented.
We additionally consider asynchronous channels with noncausal state information at both transmitter and receiver, whose causal and non-causal counterparts were analyzed in \cite{Wolfowitz1978,HeegardElGamal1983,Salehi1992,GoldsmithVaraiya1997,Rosenzweig2005}. We note that the results of this paper were partially presented in \cite{Eilat_us}.

In recent years a new technology coined as ``Cognitive Radio'' \cite{Mitola,Haykin,GoldsmithJafar2009} has emerged.
The term "cognitive radio networks" encompasses several models and definitions, however, generally speaking, the common assumption for these networks is the existence of cognitive users that can sense their surroundings and are able to change their configurations accordingly. The presence of such users in a network can drastically improve spectrum utilization and even help the non-cognitive users. In some models of cognitive radio networks, the cognitive users possess a knowledge of the codewords that licensed users transmit. Consequently, the Gel'fand-Pinsker channel, channels with side information at the transmitter and receiver, and the cognitive MAC are among the building blocks of cognitive radio networks \cite{GoldsmithJafar2009,Devroye}. The capacities of some of these synchronous channel models are known. Nevertheless, practical communication systems are not always synchronized. Examples for practical setups in which asynchronism in state information may arise:
\begin{itemize}
  \item  Multicast communication systems, in which the same message is to be transmitted to several destinations where the state sequence suffers different delays.
  \item  A communication system with no feedback, in which a cognitive transmitter obtains information about the interfering signal but does not know the time offset by which it is received since the delay towards the receiver is unknown.
  \item  A MAC with no feedback in which a cognitive user knows in advance the message which the other user (the non-cognitive user) is about to send, however, the two users may not be fully synchronized for example due to clock synchronization limitation or unknown delay in the channel.
  \item Cellular networks in which a helping interferer helps the base-stations to conceal their messages. In this setup, which is depicted in Fig.\ \ref{Eavesdroppers_fig}, several base-stations serve mobile users in the network while information leaks to the passive eavesdroppers. The helping interferer is linked to the base-stations by optical fiber channels and periodically informs them of the interfering signals it is about to transmit. Alternatively, the helping interferer and the base-stations can agree on a list of signals which the interferer will transmit in a particular order. It is also assumed that the base-stations can acquire information on the locations of users. However, synchronization issues between the helping interferer and the base-stations, the mobility of users, and unprecise users' location at the base-stations can cause the interfering signal and a base-station's transmitted signal to be out of sync. A partial list of relevant papers for the synchronous setup is \cite{Mitrpant2006,ChenVinck2008,SimeoneYener2009,KhistiDiggavi2011,Xu2014} where one can treat the side information in some of these papers as the interferer's signal.
\begin{figure}[H]
  \centering
  \includegraphics[scale=0.40]{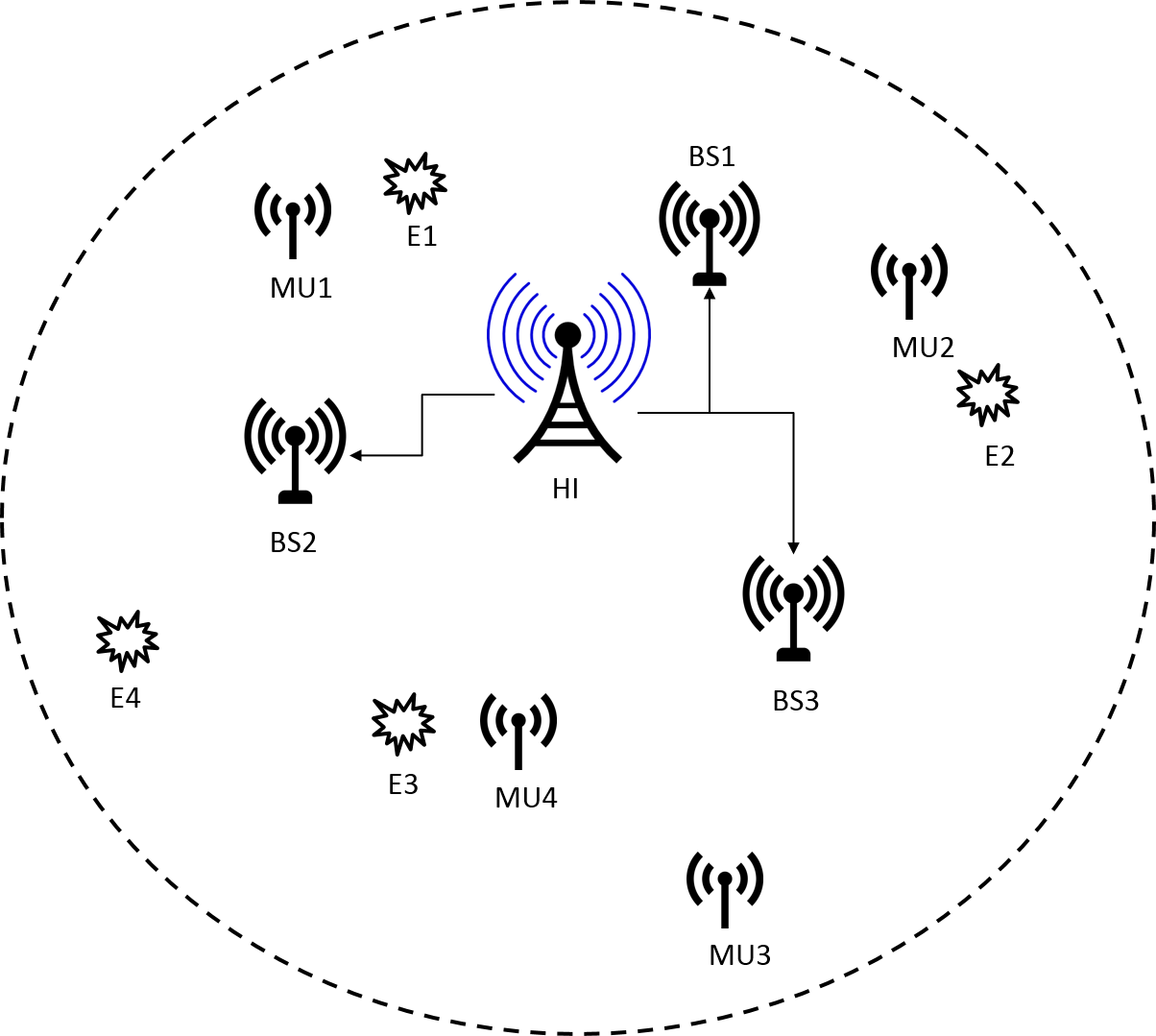}\\
  \caption{A cellular network with a helping interferer (HI), base-stations (BS), mobile users (MU) and eavesdroppers (E).}\label{Eavesdroppers_fig}
\end{figure}
\item  Cellular networks in which coordinated multipoint (CoMP) techniques are used (see for example \cite{Karakayali2006,Irmer2011}). There are several CoMP methods for the downlink which involve different schemes for cooperation and coordination of base-stations. Base stations cooperation may also occur in the uplink, for example several base-stations can jointly decode received signals. As discussed in \cite{Irmer2011}, there can be synchronization issues in these cooperative schemes. A detailed example of an asynchronous CoMP is discussed in \cite{LiuLi2013}.
\end{itemize}
We note that the results of this paper were extended to multiuser setups in \cite{us_ISIT_2014,us_paper2}.

The rest of this paper is organized as follows. In Section \ref{sec:Channel Models} we present channel models which are analyzed and define  several notations that are used throughout this paper. Subsequently, in Section \ref{sec:AGP Channel} we discuss the asynchronous Gel'fand-Pinsker channel and state lower bounds on its capacity. Section  \ref{sec:Achievable ACSI} is devoted to channels with asynchronous causal state information at the transmitter. We then present in Section \ref{sec:Achievable ACSITR} the capacity of channels with asynchronous channel state information at both the transmitter and receiver. Finally, Section \ref{sec:Conclusion} contains concluding remarks.

\section{Channel Models and Definitions}\label{sec:Channel Models}
We use the following notations and definitions:
A vector $(a_1,\ldots,a_n)$ is denoted by $a^n$, whereas the vector  $(a_i,\ldots,a_j)$ is denoted by $a_i^j$. If $a^n$ is a sequence of vectors, then the notation $a_{i,j}$ is used to address the $j$ entry of the vector $a_i$.
The probability law of a random variable $X$ is denoted by $P_X$ while $\mathcal{P}(\mathcal{X})$ denotes the set of distributions on the alphabet $\mathcal{X}$.
The set of all $n$ vectors $x^n$ that are $\epsilon$-strongly typical \cite[p.\ 326]{CT} with respect to $P_X\in {\cal P}({\cal X})$ is denoted by $T_{\epsilon}^n(X)$. Additionally, we denote  by $T_{\epsilon}^n(X|y^n)$ the set of all $n$ vectors $x^n$ that are $\epsilon$-strongly jointly typical with the vector $y^n$ with respect to a probability mass function (p.m.f.) $P_{X,Y}$.  Further, $\mathbbm{1}_{\{A\}}$ denotes the indicator function, i.e., $\mathbbm{1}_{\{A\}}$ equals $1$ if the statement $A$ holds and $0$ otherwise.

In addition, $\cal D$ is a set of integers, and $D=|\mathcal{D}|$ denotes its cardinality. Further, let $P$ be a conditional p.m.f. from $\mathcal{X}$ to $\mathcal{Y}$. For $x^D\in\mathcal{X}^D$ denote by $\{P_d(y|x_1^D)\}$ a set of conditional p.m.f.'s from ${\cal X} ^D$ to $\cal Y$,  that depend on the value of $d$, where $d\in \mathcal{D}$.
We use the notation $T_{d,\epsilon}^n(X,Y)$  to make the underlying p.m.f. $P_d(x_1^D,y)$ explicit where $d\in\cal D$. Similarly, we use the notation $T_{p,\epsilon}^n(X)$  to make the underlying p.m.f. $p$ explicit.

We next describe the channel models of the aforementioned channels.

\subsection {The Asynchronous Gel'fand-Pinsker Channel}
\begin{figure}[H]
\centering	
\includegraphics{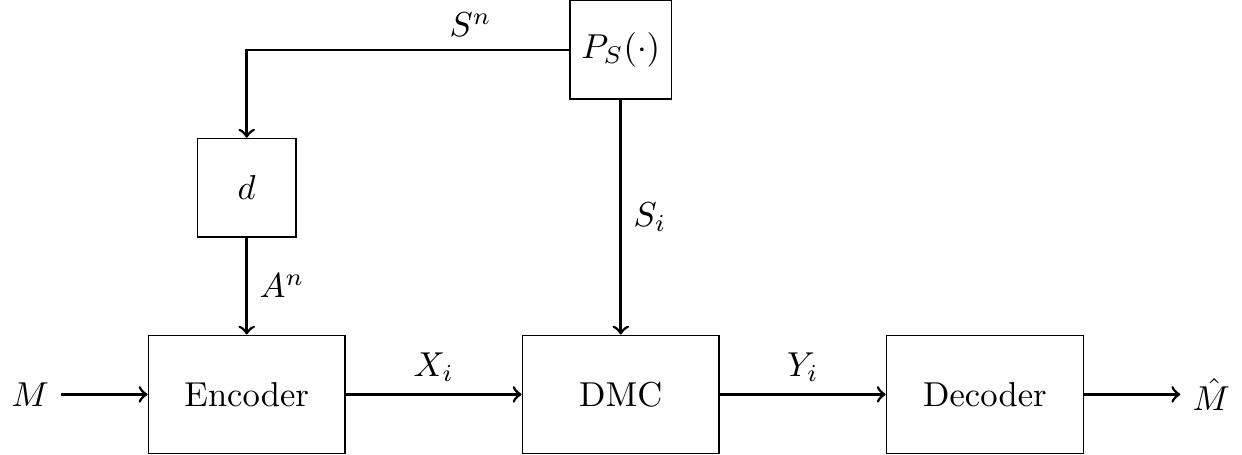}
\caption{Asynchronous Gel'fand-Pinsker channel.}
\label{AGP_fig}
\end{figure}

The asynchronous Gel'fand-Pinsker channel (AGP channel), which is depicted in Fig.\ \ref{AGP_fig}, is a discrete memoryless stationary and state-dependent channel.
It is defined by the channel transition probabilities $\left\{P(y|x,s)\right\}$, the channel input alphabet $\cal X$, the channel output alphabet $\cal Y$, the state symbol alphabet $\cal S$, and the state sequence distribution, which is assumed to be i.i.d. $P_S$. The transmitter observes non-causally a possibly delayed version of the states sequence $(S_1,\ldots,S_n)$. In other words, before the beginning of transmission, the transmitter observes a sequence $(A_1,\ldots,A_n)$ of state symbols according to:
\begin{flalign}\label{equation1}
(A_1,\ldots,A_n)=
\begin{cases}
   \left(Z_1,\ldots,Z_{d},S_1,S_{2},\ldots,S_{n-d}\right), & \text{if } d\geq 0 \\
   \left(S_{1-d},\ldots,S_n,Z_1,\ldots,Z_{-d}\right),       & \text{if } d<0
  \end{cases}
  \end{flalign}
where $d\in  \mathcal{D}$, and $Z_1,Z_2,\ldots,Z_{d}$ are i.i.d. with $Z_i\sim P_S$ independent of $(S_1,\ldots,S_n)$. Since $A^n$ is a possibly delayed version of the sequence $S^n$, it follows that $\mathcal{A}=\mathcal{S}$ and $A^n\in\mathcal{S}^n$.

Let $x^n\in\mathcal{X}$ and $s^n\in\mathcal{S}^n$ be the codeword and the state-sequence, respectively, and let
$y^n\in\mathcal{Y}^n$ be the output of the channel.
The conditional distribution of $Y^n$ given $(X^n,S^n)$ is given by
\begin{flalign}\label{trans_prob_ACMAC}
P(y^n |x^n,s^n)=\prod_{i=1}^n P_{Y|X,S}(y_i|x_i,s_i).
\end{flalign}

Let $\mathcal{M}=\{1,2,\ldots,2^{nR}\}$, and assume that the massage $M$ is a random variable uniformly distributed over the set $\mathcal{M}$.
A $(2^{nR},n)$-code for the AGP channel consists of an encoding function
\begin{flalign}
&f_n: \mathcal{M} \times \mathcal{S}^n\rightarrow\mathcal{X}^n
\end{flalign}
and a decoding function
\begin{flalign}
&g_n:\mathcal{Y}^n \rightarrow \mathcal{M}.
\end{flalign}
Define the average probability of error for $d\in \mathcal{D}$ as
\begin{flalign}
\bar{P}_{e,d} =  \frac{1}{2^{nR}} \sum_{m=1}^{2^{nR}} &\sum_{\substack{(s^n ,a^n) \in \mathcal{S}^n \times \mathcal{S}^n, \\ y^n: g_n(y^n)\neq m}}
 P_d(s^n,a^n)P\left(y^n |f_n(m,a^n),s^n \right),
\end{flalign}
where
\begin{flalign}\label{P_dsa}
P_d(s^n,a^n) =
  \begin{cases}
  P(s^n)P(a_{n+d+1}^n)\cdot\mathbbm{1}_{\left\{s_{1-d}^n=a_1^{n+d}\right\}},       & \text{if } d<0\\
   P(s^n)\cdot\mathbbm{1}_{\left\{s^n=a^n\right\}}, & \text{if } d=0 \\
   P(s^n)P(a_1^d)\cdot\mathbbm{1}_{\left\{s_1^{n-d}=a_{d+1}^n\right\}},       & \text{if } d>0
  \end{cases}
\end{flalign}
$P(s^n)=\prod_{i=1}^n P_S(s_i)$ and $P(a^n)=\prod_{i=1}^n P_S(a_i)$.

A $(2^{nR},n)$-code for the AGP channel is said to be a $(2^{nR},n,\epsilon)$-code if $\bar{P}_{e,d}\leq\epsilon$ for all $d \in \mathcal{D}$.
A rate $R$ is said to be achievable for the AGP channel, if there exists a sequence of $\left(2^{nR},n,\epsilon_n \right) $-codes with $\epsilon_n \rightarrow 0$ as $n \rightarrow \infty$.
\newline
The capacity of the AGP channel, $C_{AGP}$, is the supremum of all achievable rates.

\subsection {The Causal Case} \label{acsi_def}
We next introduce a state dependent channel with asynchronous causal\footnote{We refer to this setup as the causal case, but in fact, if $d<0$, the cognitive user has a lookahead of $d$ future symbols.} state information (ACSI) at the transmitter. We refer to this channel as the ACSI channel.

The definitions for the ACSI channel are similar to those of the AGP channel, with the following modifications:

In this setup, before transmitting $X_i$, the encoder observes $A_1,\ldots,A_i$ which are defined in (\ref{equation1}) (rather than $A_1,\ldots,A_n$).

As before, it is assumed that the messages are equiprobable over $\mathcal{M}$. A $(2^{nR},n)$-code for the ACSI channel consists of the encoding functions $\{f_i\}, i=1,\ldots,n$ where
\begin{equation}
f_i: \mathcal{M}\times {\cal S}^i\rightarrow {\cal X}
\end{equation}
and a decoding function
\begin{flalign}
&g_n:\mathcal{Y}^n \rightarrow \mathcal{M}.
\end{flalign}

The average probability of error is given by,
\begin{flalign}
\bar{P}_{e,d} =  \frac{1}{2^{nR}} & \sum_{m=1}^{2^{nR}} \sum_{\substack{(s^n ,a^n) \in \mathcal{S}^n \times \mathcal{S}^n,\\y^n: g_n(y^n)\neq m}}
 P_d(s^n,a^n)\prod_{i=1}^nP\left(y_i |f_i(m,a^i),s_i \right),
\end{flalign}
where $P_d(s^n,a^n)$ is defined in (\ref{P_dsa}).
\newline
The definitions of the achievable rate and the capacity are similar to those of the AGP channel.

\subsection {Asynchronous Channels with States Available Non-Causally Both at the Transmitter and Receiver}\label{ACSITR_def}
\begin{figure}[H]
\centering	
\includegraphics{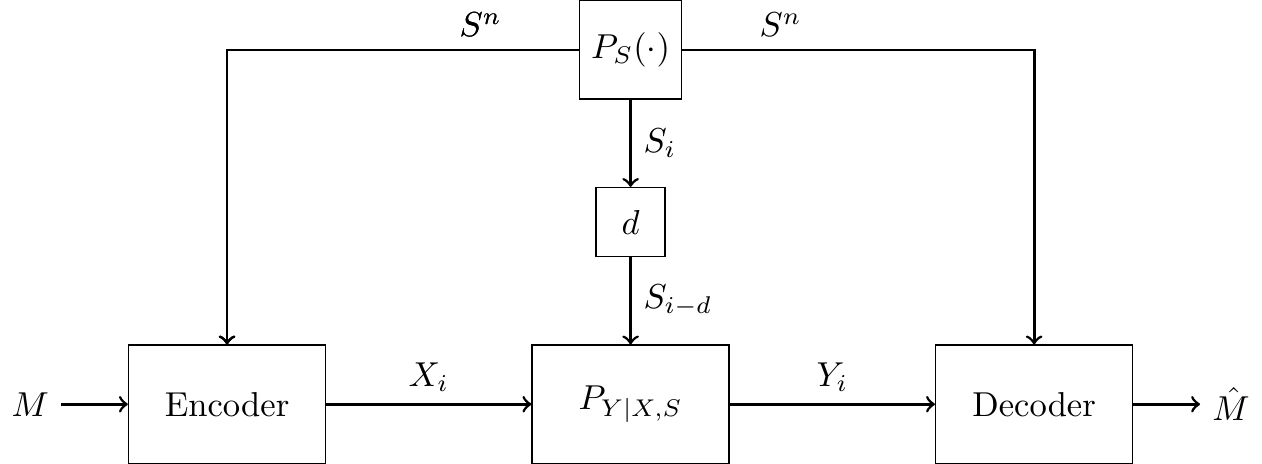}
\caption{{\footnotesize Asynchronous Channels with States Non-causally Available both at the Transmitter and the Receiver.}}
\label{Delay_trans_receiv_fig}
\end{figure}

An asynchronous channel with channel states non-causally known  at both the transmitter and receiver (see Fig.\ \ref{Delay_trans_receiv_fig}) is a stationary discrete memoryless state-dependent channel, defined by $\{P(y|s,x)\}, \cal X,\cal Y,\cal S$, and $P_S$ as before. Both the transmitter and the receiver observe non-causally the sequence $(S_1,\ldots,S_n)$, and in addition the link between the state source and the channel may suffer a delay $d$
where $d\in\mathcal{D}$.

Let the random message $M$ be defined as before, i.e., distributed equiprobably over ${\cal M}$.
A $(2^{nR},n)$-code for the asynchronous channel with channel states non-causally known both at the transmitter and receiver, consists of an encoding function
\begin{flalign}
&f_n:\mathcal{M} \times \mathcal{S}^n\rightarrow\mathcal{X}^n \end{flalign}
and a decoding function
\begin{flalign}
&g_n:\mathcal{Y}^n\times \mathcal{S}^n \rightarrow \mathcal{M}.
\end{flalign}
Define the average probability of error for $d\in\mathcal{D}$ as
\begin{flalign}
\bar{P}_{e,d} =  \frac{1}{2^{nR}} \sum_{m=1}^{2^{nR}} & \sum_{\substack{s^n \in \mathcal{S}^n,\\y^n: g_n(y^n,s^n)\neq m}} P(s^n)P_d\left(y^n |f_n(m,s^n),s^n \right),
\end{flalign}
$P(s^n)=\prod_{i=1}^n P_S(s_i)$ and
\begin{flalign}
P_d(y^n|x^n,s^n)=\prod_{i=1}^n P(y_i|x_i,s_{i-d}),
\end{flalign}
where for all $i\in \{1,\ldots,n\}$ such that $i-d\notin\{1,\ldots,n\}$,  $s_{i-d}$ are arbitrary.

A $(2^{nR},n)$-code is said to be a $(2^{nR},n,\epsilon)$-code if $\bar{P}_{e,d}\leq\epsilon$ for all $d \in \mathcal{D}$.
A rate R is said to be achievable for the asynchronous channel with channel states non-causally known both at the transmitter and receiver, if there exists a sequence of $\left(2^{nR},n,\epsilon_n \right) $-codes with $\epsilon_n \rightarrow 0$ as $n \rightarrow \infty$.

The capacity of the asynchronous channel with channel states non-causally known both at the transmitter and receiver, $C_{ACSITR}$, is the supremum of all achievable rates.

\subsection{The Set of Possible Delays}
For simplicity of the presentation, throughout this paper, we assume that the set of possible delays in the aforementioned channels is
$\mathcal{D}=\{-d_{min},-d_{min}+1,\ldots,d_{max}\}$, where $0 \leq d_{min},d_{max}$, it follows that $D=d_{max}+d_{min}+1$. Additionally, throughout this paper we assume that all transmitters and receivers know a-priori the (finite) values $d_{min}$ and $d_{max}$. We note that the results which are derived in this paper can be easily generalized to arbitrary finite sets of delays, and hold in the general case in which the delay is randomly distributed over a finite set.

\subsection{Known Delay at the Receiver}
In all of the above channel models, i.e., the AGP, ACSI and the asynchronous channel with channel states non-causally known at both the transmitter and receiver, we assume that the decoder does not know the actual delay in the channel before decoding the message. However, since the set of delays $\mathcal{D}$ is finite, by sending predefined training sequences in the first $o(n)$ bits, the decoder can deduce the delay with probability of error that vanishes as $n$ tends to infinity. Therefore, we can assume hereafter that the decoder knows the delay $d$ prior to the decoding stage. We will however include transmission of the training sequence in our coding schemes.

\section{The AGP Channel}\label{sec:AGP Channel}
In this section we derive lower bounds for the capacity of the AGP channel, when the alphabets $\mathcal{X},\mathcal{S},\mathcal{Y}$, and the delay set, ${\cal D}$ are finite. In addition, we state the capacity of the AGP channel with feedback for finite delays.
\subsection{An Achievable Rate for the AGP Channel}\label{sec:Achievable AGP}
The single-letter formula for the capacity of the synchronous Gel'fand-Pinsker (GP) channel $P_{Y|X,S}$ is given by \cite{GP}
\begin{equation}\label{eq: GP channel capacity}
C_{GP}=\max_{P_{U,X|S}}\left[I(U;Y)-I(U;S)\right]
\end{equation}
where $U-(S,X)-Y$ is a Markov chain, and $|\mathcal{U}|\leq|\mathcal{X}|\cdot |\mathcal{S}|$.

We next present an achievable rate for the AGP channel. In Section \ref{sec:Example} we prove that this lower bound is tight for the binary symmetric AGP channel with crossover probability of $0.5$ and $\mathcal{D}=\{0,1\}$.

\begin{theorem}\label{theorem_AGP1}
The rate
\begin{align} \label{AGP_rate1}
	R&=\max _ {p_{U,X|A}} \left[ \frac{1}{D} \cdot I_{p_1}(U;Y)
	+\frac{D-1}{D}\cdot I_{p_2}(U;Y)- I(U;A)\right]
\end{align}
where,
\begin{align} \label{eq:prob}
	&p_1(u,y) = \sum_{a,x} P_S(a)P_{U,X|A}(u,x|a)P_{Y|X,S}(y|x,a) \nonumber\\
	&p_2(u,y) =  \sum_{s,a,x} P_A(a)P_{U,X|A}(u,x|a)P_S(s)P_{Y|X,S}(y|x,s)\nonumber\\
	&p_{U,A}(u,a) =  \sum_x P_A(a) P_{U,X|A}(u,x|a) \nonumber \\
	&P_A(a) =P_S(a) \quad \forall a \in \cal S
\end{align}
is achievable for the AGP channel with channel conditional distribution $P_{Y|X,S}$ and a set of delays $\mathcal{D}$.
\end{theorem}
Note that the rate in (\ref{AGP_rate1}) converges to the channel capacity with no side information, as $D$, the size of the set of all possible delays, tends to infinity. In addition, if $d_{min}=d_{max}=0$, that is, there is no actual delay in the channel, the channel degenerates to the Gelf'and-Pinsker channel, as the formula (\ref{AGP_rate1}) indicates.

The coding scheme that is used in the proof employs binning and "segment time sharing". In segment time sharing, the codeword is partitioned to several segments. In each segment the encoder chooses a different encoding function (similarly to the ordinary time sharing). To decode the message, the decoder which knows the identity of the segments \emph{jointly decodes} the segments. That is, unlike the ordinary time sharing, the decoder in segment time sharing jointly decodes all the segments:
The main idea of the proof is that the encoder uses the GP coding scheme for each possible delay by dividing the codeword into equal length segments. In each of these segments the encoder assumes a different delay (out of the set $\mathcal{D}$). The decoder knows for each segment the assumed delay which was decided by the encoder. Additionally, as mentioned before, we can assume that the decoder knows the actual delay of the side information (for example by sending a training sequence). Knowing this delay the decoder looks for a codeword such that each of its segments is typical with its corresponding output according to the p.m.f which is induced by the channel transition probability and the assumed delay of the segment in the encoding stage. In the AGP setup segment time sharing yields better results than ordinary time sharing since the redundancy in one segment can help in decoding another segment.
For the detailed proof see Appendix \ref{AGP_A1}.

\subsection{An Example - The Binary Symmetric AGP Channel} \label{sec:Example}
Consider the binary symmetric AGP (BS-AGP) channel  defined by the input-output relation,
\begin{flalign}
Y_i=X_i\oplus S_i
\end{flalign} where $S_i \sim \text{Bernoulli}\left(\frac{1}{2} \right)$, and with $d\in\{0,1\}$. \newline
In the ordinary synchronous GP setup, a capacity achieving scheme is to construct a codebook containing all the possible binary vectors $u^n\in \{0,1\}^n$. To transmit the vector $u^n$, the transmitter sends $x_i=u_i\oplus s_i$. Consequently, the received $i$th symbol is $y_i=u_i$, and the resulting achievable rate is thus that of the clean channel $y_i=u_i$, i.e., 1 bit per channel use.
In the asynchronous case, consider the following coding scheme which is a special case of the general scheme presented in Section \ref{sec:Achievable AGP}. A codebook containing $2^{\frac{n}{2}-1}$ binary codewords $u^n$ with binary $\text{Bernoulli}\left(\frac{1}{2}\right)$ symbols is drawn. Recall that $A_1,\ldots,A_n$ is the (possibly delayed) observed state sequence and let
\begin{flalign} \label{coding_scheme_2_delays}
x_i =
  \begin{cases}
   u_i\oplus a_i, &  i \in \left\{1,\ldots,\frac{n}{2}\right\} \\
   u_i\oplus a_{i+1},       &  i \in \left\{\frac{n}{2}+1,\ldots,n-1\right\}\\
   u_i, & \text{for } i=n
  \end{cases}.
\end{flalign}
Let $P_{UY}$ be the product p.m.f., i.e. $P_{U,Y}(u,y)=P_U(u)P_Y(y)$. The decoder looks for a sequence $u^n$ such that
\begin{flalign}
	 u_1^{\frac{n}{2}} = y_1^{\frac{n}{2}}
	\text{ and }\left( u_{\frac{n}{2}+1}^{n-1} , y_{\frac{n}{2}+1}^{n-1} \right)\in T_{P_{UY},\epsilon}^{\frac{n}{2}-1}(U,Y)
\end{flalign} if $d=0$,
or such that
\begin{flalign}
	u_{\frac{n}{2}+1}^{n-1} = y_{\frac{n}{2}+1}^{n-1}
\text{ and }\left( u_1^{\frac{n}{2}} , y_1^{\frac{n}{2}} \right)\in T_{P_{UY},\epsilon}^{\frac{n}{2}}(U,Y).
\end{flalign}
if $d=1$.
In the case of no delay, $d=0$, this results in,
\begin{flalign}
y_i =x_i \oplus s_i =
    \begin{cases}
   u_i, &  i \in \left\{1,\ldots,\frac{n}{2}\right\} \\
   u_i \oplus s_{i+1} \oplus s_i,   &  i \in \left\{\frac{n}{2}+1,\ldots,n\right\}\\
   u_i  \oplus s_i, & i=n
  \end{cases}.
\end{flalign}

In the case where $d=1$, this results similarly in
\begin{flalign}
y_i =x_i \oplus s_i &=
    \begin{cases}
   u_i, &  i \in \left\{1,\ldots,\frac{n}{2}\right\} \\
   u_i\oplus s_{i-1} \oplus s_i,   &  i \in \left\{\frac{n}{2}+1,\ldots,n\right\}\\
   u_i  \oplus s_i, & i=n
  \end{cases}.
\end{flalign}
Define the random variables $K_i = S_{i-1} \oplus S_i$ where $K_i \sim \text{Bernoulli} \left(\frac{1}{2}\right)$, and the random variables $L_i = S_{i+1} \oplus S_i$ where $L_i \sim \text{Bernoulli} \left(\frac{1}{2}\right)$. Clearly, this scheme can guarantee reliable decoding of the message for all rates lower than
\begin{flalign}\label{BSC_upper_rate}
 R_l(0.5)&= \frac{1}{2}I(U;U\oplus L)+\frac{1}{2}I(U;U) -I(U;A)\nonumber\\
 &=\frac{1}{2}I(U;U)+\frac{1}{2}I(U;U\oplus K) -I(U;A)\nonumber\\
 &=\frac{1}{2}\cdot 1 +\frac{1}{2}\cdot 0 - 0 =\frac{1}{2}.
\end{flalign}
In the general case, i.e. the BS-AGP channel with crossover probability $p$, a similar coding scheme assures reliable decoding of the message for all rates lower than
\begin{flalign}\label{BSC_maximal_rate}
 R_l(p)=\frac{1}{2}+\frac{1}{2}\left[1-h_2\left(2p(1-p)\right)\right],
\end{flalign}
where $h_2(x)=-x\log_2(x)-(1-x)\log_2(1-x)$.

\begin{figure}[H]
\centering
	\includegraphics[scale=0.45]{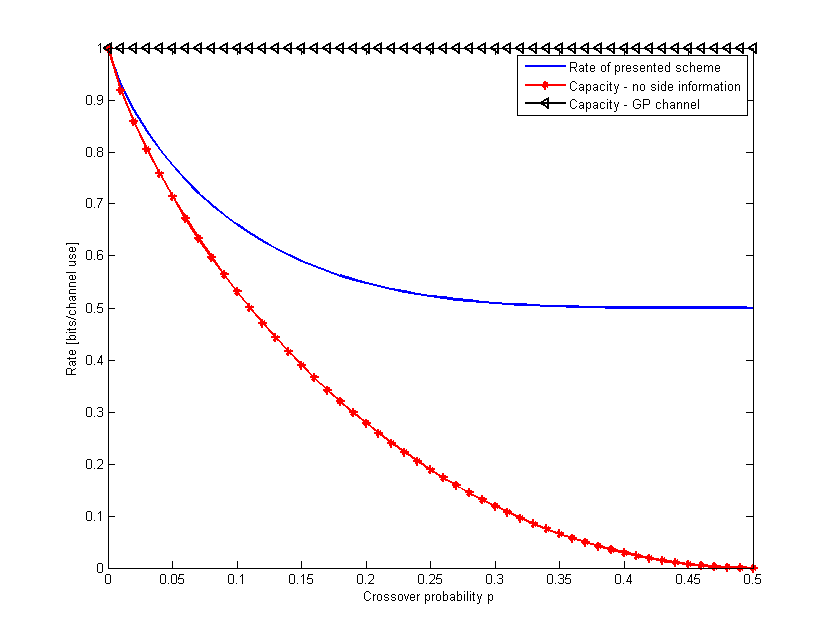}
	\caption{Comparison of the achievable rates for the BS-AGP channel with crossover probability $p$ and $\mathcal{D}=\{0,1\}$, the capacity with no side information, and the capacity of the synchronous Gel'fand-Pinsker channel.}
\label{BSC_example_fig}
\end{figure}

In Fig.\ \ref{BSC_example_fig}, we compare the lower bound $R_l(p)$ to the capacity of the binary symmetric Gel'fand-Pinker channel and to the capacity of the binary symmetric channel (BSC) with no side information at the encoder nor the decoder, all channels have crossover probability $p$.

An important question is whether the rate $R_l(p)$ is the capacity of the  BS-AGP channel with crossover probability of $p$ and whether the answer depends on the crossover probability $p$?

To answer these questions we rely on a relevant setup which is considered in \cite{Khisti}. The state dependent binary multicast channel that is studied in \cite{Khisti} is composed of the input sequence $X^n$, the channel states sequence $\{S_{(1)}^n,S_{(2)}^n\}$ and the output sequences. The state dependent binary multicast channel is defined by the following input-outputs relations,
\begin{flalign}
Y_{(k)}^n=X^n\oplus S_{(k)}^n,\quad k\in\{1,2\}
\end{flalign}
where $S_{(1)}^n,S_{(2)}^n,X^n,Y_{(1)}^n,Y_{(2)}^n\in\{0,1\}^n$, and $\oplus$ is a symbol-by-symbol modulo-$2$ operation.
\newline
It is known \cite{Khisti}, that for two correlated sequences $S_{(1)}^n,S_{(2)}^n$ which are not necessarily i.i.d. processes, the capacity of the binary multicast  channel is,
\begin{flalign}\label{BSC_capacity}
C = 1-\frac{1}{2} \lim_{n\rightarrow\infty}\frac{1}{n}H(S_{(1)}^n\oplus S_{(2)}^n).
\end{flalign}

The BS-AGP channel can be identified with the binary multicast channel that appears in \cite{Khisti}, where  $S_i$ and $S_{i-1}$ play the roles of $S_{1,i}$ and $S_{2,i}$, respectively. Note that the process $\{(S_i,S_{i-1})\},\hspace{0.2cm}i=1,\ldots,n$ is not an i.i.d. process.

Now, knowing the capacity of the channel, we can conclude that the  capacity of the BS-AGP channel with crossover probability $\frac{1}{2}$ is a special case of (\ref{BSC_capacity}). This is true since the process $\{S_i\oplus S_{i-1}\}, i=1,2,\ldots $ is i.i.d. when $p=\frac{1}{2}$. However, if $p\notin \{0,0.5,1\}$, then $S_1^n\oplus S_0^{n-1}$ is not a memoryless or constant sequence. Therefore,
\begin{flalign}\label{lower_bound_capacity}
C &= 1-\frac{1}{2} \lim_{n\rightarrow\infty}\frac{1}{n}H(S_1^n\oplus S_0^{n-1}) \nonumber\\
&> 1-\frac{1}{2} \lim_{n\rightarrow\infty}\frac{1}{n}\sum_{i=1}^n H(S_i \oplus S_{i-1})\nonumber\\
&= 1-\frac{1}{2} H(S_1 \oplus S_2) = 1-\frac{1}{2} h_2(p(1-p))\triangleq R_l(p).
\end{flalign}
That is,  $R_l(p)$ is not the capacity of the BS-AGP channel when $p\notin \{0,0.5,1\}$.
Another insight from equation (\ref{BSC_capacity}) is that in some channels there is a gain in using multi-letters coding. This insight will be used later in our generalized scheme.

\begin{figure}[H]
\centering
	\includegraphics[scale=0.45]{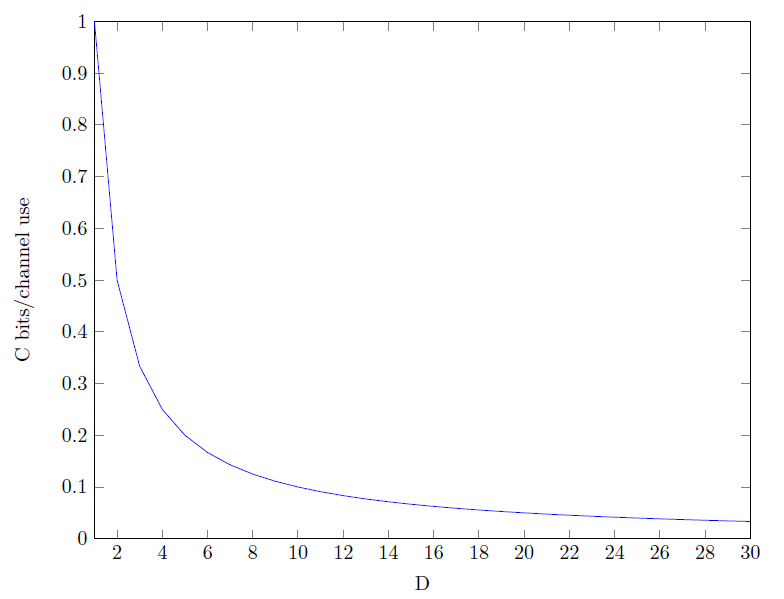}
	\caption{A lower bound on the capacity of the BS-AGP channel with crossover probability $\frac{1}{2}$ as function of the size of the set of all possible delays.}
\label{BSC_example_several_delays}
\end{figure}

An additional issue concerns the usefulness of the side information for the BS-AGP. For simplicity we consider the  BS-AGP channel with crossover probability $0.5$ with different cardinality of the delay set $\mathcal{D}$. As mentioned before for $p=\frac{1}{2}$, the sequence $\{(S_i\oplus S_{i-d})\},\hspace{0.2cm}i=1,\ldots,n$  is i.i.d. with p.m.f. $\text{Bernoulli}\left(\frac{1}{2}\right)$ for each $d\in \mathcal{D}$. By generalizing the coding scheme in Eq. (\ref{coding_scheme_2_delays}) for the set of delays of cardinality $D$ and by Theorem \ref{theorem_AGP1}, all rates which are not greater than $\frac{1}{D}$ bits/channel use are achievable. Fig.\ \ref{BSC_example_several_delays} depicts the lower bound on the capacity of the BS-AGP channel with crossover probability $0.5$ with respect to the number of possible delays $D$.

\subsection{An Improved Lower bound for the AGP Channel}
For $d_{min},d_{max}<\infty$ we can identify the AGP channel with a multicast channel with $D=d_{max}+d_{min}+1$ users, in which all the users share the same channel transition probabilities, but differ in the fact that the state of channel $k\in\{1,\ldots,D\}$ at time $i$ is $S_{i-d_{min}+k-1}$.
\begin{figure}[H]
\centering	
\includegraphics{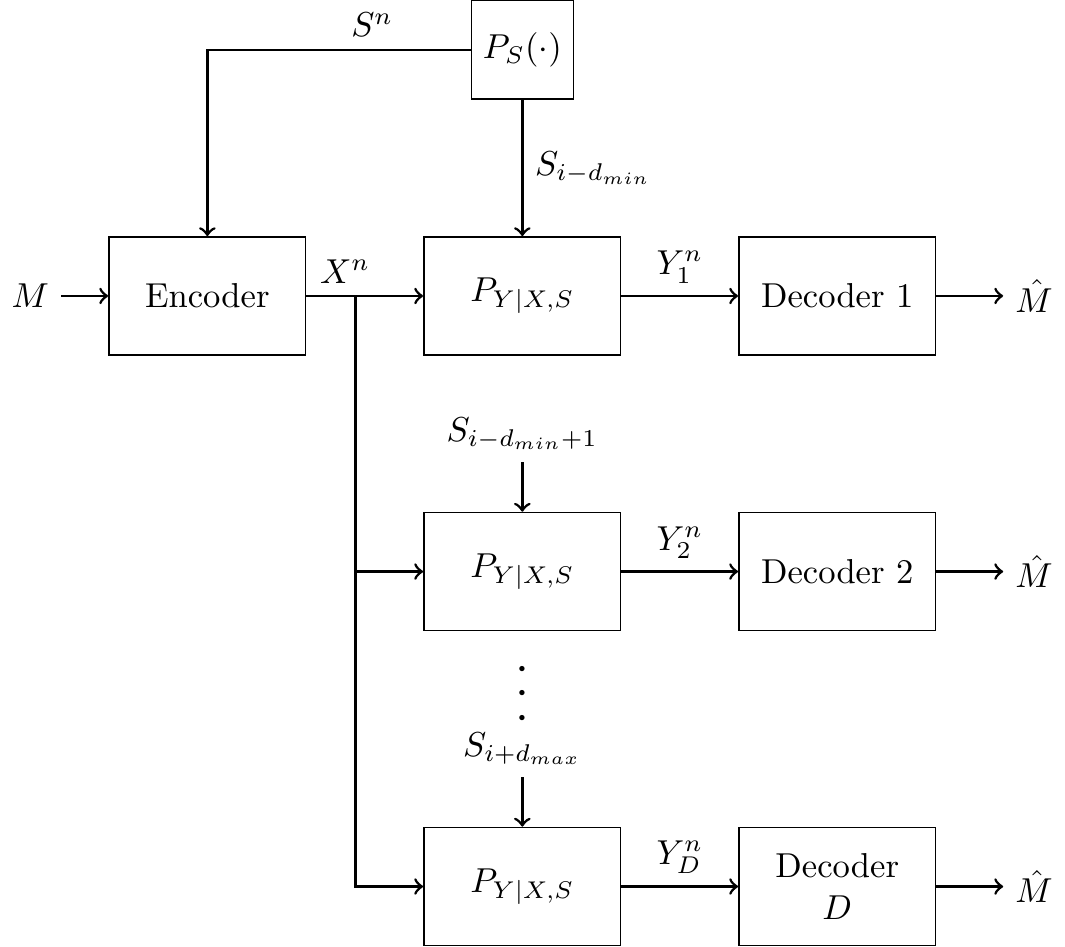}
\caption{Multicast channel interpretation.}
\label{AGP_multicast_fig}
\end{figure}

\begin{figure}[H]
\centering	
\includegraphics{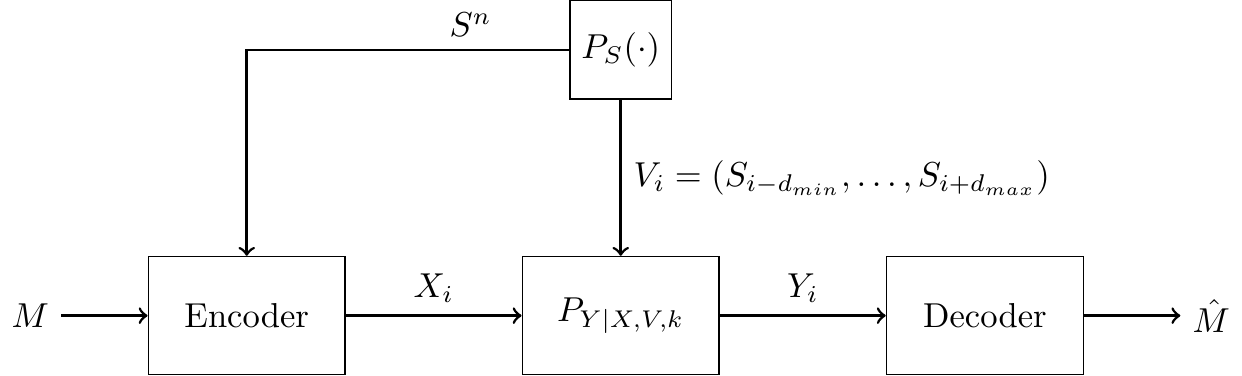}
\caption{Compound channel interpretation.}
\label{AGP_compound_fig}
\end{figure}

In addition, similarly to \cite{Piantanida,Nair}, the channel in Fig.\  \ref{AGP_multicast_fig} has a compound channel representation depicted in Fig.\ \ref{AGP_compound_fig} where $P(y|x,v,k)=P_{Y|X,S}(y|x,v_k)$,  $k\in\{1,\ldots,D\}$, and $V_i=(S_{i-d_{min}},\ldots,S_{i+d_{max}})$
is the vector of all possible channel states at time $i$.

We note that the $D-$tuples: $V_1^n=(S_{i-d_{min}},\ldots,S_{i+d_{max}})_{i=1}^n$ are statistically dependent, additionally
\begin{flalign}\label{Pv1}
P_{V_i}(v)=P_V(v)\triangleq \prod_{\ell=1}^{D}P_S(v_{\ell}),
\end{flalign}
where $v=(v_1,\ldots,v_D)$.

We next present improved achievable rates for the AGP channel. The coding schemes that achieve these
rates are extensions of Theorems 2.4 and 2.6 in \cite{Piantanida} and of Theorem 1 in \cite{Nair}.

For the sake of clarity, we first present an achievable rate for the case  $d\in\{0,1\}$.

\begin{theorem}\label{theorem_AGP2}
Let $P_{Y|X,S}$ be the channel conditional distribution of an AGP channel with a set of delay $\mathcal{D}=\{0,1\}$. The rate,
\begin{flalign}\label{AGP_lower_rate}
R=\max_{P_T,P_{W,U_1,U_2,X|V,T}}\min\left\{\right.&I(W,U_1;Y_1|T)-I(W,U_1;V|T), \nonumber\\
&I(W,U_2;Y_2|T)-I(W,U_2;V|T),\nonumber\\
&\frac{1}{2}\left[I(W,U_1;Y_1|T)-I(W,U_1;V|T)\right.\nonumber\\
&\left.\quad+I(W,U_2;Y_2|T)-I(W,U_2;V|T)-I(U_1;U_2|W,V,T)\right]\left.\right\}
\end{flalign}
is achievable where
$V=(V_1,V_2)$ and $P_V(v_1,v_2)=P_S(v_1)P_S(v_2)$, and
\begin{flalign}\label{probability_theorem_AGP2}
&P_{T,V,W,U_1,U_2,X,Y_k}=P_TP_{V}P_{W,U_1,U_2,X|V,T}P_{Y=Y_k|X,S=V_k},\quad k\in\{1,2\}.
\end{flalign}
\end{theorem}
This result can be generalized for any $0\leq d_{min},d_{max}<\infty$.
\begin{theorem}\label{Theorem_AGP_general_rate}
Let $P_{Y|X,S}$ be the channel conditional distribution of an AGP channel with a set of delay $\mathcal{D}=\{-d_{min},\ldots,d_{max}\}$. Denote
 $\mathcal{D}_1=\{1,\ldots,D\}$ where $D=d_{min}+d_{max}+1$.
The rate,
\begin{flalign}\label{AGP_general_rate}
&R=\max_{P_T,P_{W,U_1,\ldots,U_{D},X|V,T}}\min_{\mathcal{L} \subseteq \mathcal{D}_1}\left\{\frac{1}{\left\|\mathcal{L}\right\|}\left[\sum_{l\in\mathcal{L}}
I(W,U_l;Y_l|T)-\left\|\mathcal{L}\right\|\cdot I(W;V|T)\right.\right.\nonumber\\
&\hspace{6.2cm}\left.\left.+H(\{U_{\ell}|\ell\in\mathcal{L}\}|W,V,T)-\sum_{l\in\mathcal{L}} H(U_l|W,T)\right]\right\}
\end{flalign}
is achievable for the AGP channel, where
$P_V$ is given in (\ref{Pv1}), and
\begin{flalign}
&P_{T,V,W,U_1,\ldots,U_D,X,Y_k}=P_TP_VP_{W,U_1,\ldots,U_D,X|V,T}P_{Y=Y_k|X,S=V_k},\enskip\forall k\in\mathcal{D}_1.
\end{flalign}
\end{theorem}

For the simplicity of the presentation we prove only Theorem \ref{theorem_AGP2}, the proof appears in Appendix \ref{AGP_A2}. The proof of Theorem \ref{Theorem_AGP_general_rate} consists of similar steps and therefore is omitted.

In addition, in light of the BS-AGP channel example, a coding scheme which involves multi-letter coding can achieve higher rates. Therefore, we include multi-letter coding in the coding scheme of Theorem \ref{Theorem_AGP_general_rate}. This yields the following result.

\begin{theorem}\label{theorem_AGP3}
Let $P_{Y|X,S}$ be the channel conditional distribution of an AGP channel with a set of delays $\mathcal{D}=\{-d_{min},\ldots,d_{max}\}$. Denote
 $\mathcal{D}_1=\{1,\ldots,D\}$ where $D=d_{min}+d_{max}+1$.
The rate,
\begin{flalign}
&R=\lim_{n\rightarrow \infty}\sup_{P_T,P_{W^n,U_1^n,\ldots,U_{D}^n,X^n|V^n,T}}\min_{\mathcal{L} \subseteq \mathcal{D}_1}
\quad\left\{\frac{1}{n}\cdot\frac{1}{\left\|\mathcal{L}\right\|}\left[\sum_{l\in\mathcal{L}}
I(W^n,U_l^n;Y_l^n|T)-\left\|\mathcal{L}\right\|\cdot I(W^n;V^n|T)\right.\right.\nonumber\\
&\hspace{6cm}\left.\left.+H(\{U_{\ell}^n|\ell\in\mathcal{L}\}|W^n,V^n,T)-\sum_{l\in\mathcal{L}} H(U_l^n|W^n,T)\right]\right\}
\end{flalign}
is achievable for the AGP channel, where
$P_V^n=\prod_{i=1}^nP(v_i|v_{i-1})$,
\begin{flalign}
&P(v_1)=\prod_{j=1}^{D}P_S(v_{1,j}), \label{v_memoryless_source1}\\
&P(v_i|v_{i-1})= \mathbbm{1}_{\{(v_{i,1},\ldots,v_{i,D-1})=(v_{i-1,2},\ldots,v_{i-1,D})\}}P_S(v_{i,D}),\quad 2\leq i\leq n,\label{v_memoryless_source2}
\end{flalign}
and $v_{i,j}$ denotes the $j$-th entry in of the vector $v_i$.
Additionally,
\begin{flalign}
&P_{T,V^n,W^n,U_1^n,\ldots,U_D^n,X^n,Y_k^n}=
P_TP_{V^n}P_{W^n,U_1^n,\ldots,U_D^n,X^n|V^n,T}\prod_{i=1}^nP_{Y=Y_{k,i}|X=X_i,S=V_{k,i}},\enskip\forall k\in\mathcal{D}_1.
\end{flalign}
\end{theorem}

Finally, we remark that similar results hold for stationary Markov state-source, with the exception that equations  (\ref{Pv1}) and (\ref{v_memoryless_source1})-(\ref{v_memoryless_source2}) are replaced with the probability law of the Markov source. Furthermore, this is also true for stationary and ergodic state-source, where equations (\ref{Pv1}) and (\ref{v_memoryless_source1})-(\ref{v_memoryless_source2}) are changed according to the state-source distribution.

\subsection{The AGP Channel with Feedback} \label{sec:Feedback}

In this section we consider the AGP channel with feedback. In this setup, in addition to the non-causal knowledge of $A^n$, at each time instant $i$ the encoder observes $Y^{i-1}$. We assume that $d_{max},d_{min}<\infty$ and, as before, the delay is fixed throughout the transmission of a codeword. Unlike the case of the AGP channel, in this case the encoder can recover the actual delay of the side information by sending a training sequence. We next prove that the capacity of the AGP channel with feedback is equal to the capacity of the GP channel.
\begin{theorem}
The capacity of the AGP channel with feedback is given by
\begin{equation}
C_{GP}=\max_{P_{U,X|S}}\left[I(U;Y)-I(U;S)\right].
\end{equation}
\end{theorem}
\begin{IEEEproof}
It is known \cite{Merhav} that the capacity of the GP channel with feedback is equal to $C_{GP}$ - the capacity of the channel without feedback. To achieve $C_{GP}$, the encoder first transmits a training sequence of length $T(n)$ such that $\lim_{n\rightarrow\infty} T(n)=\infty$ and $T(n)=o(n)$, designated to inform the transmitter of the delay via feedback. Once the delay is recovered by the encoder, an ordinary GP coding scheme can  be applied in the remaining $n-T(n)$ channel uses.
\end{IEEEproof}

{
   \def\OldComma{,}
    \catcode`\,=13
    \def,{%
      \ifmmode%
        \OldComma\discretionary{}{}{}%
      \else%
        \OldComma%
      \fi%
    }%
\section{An Achievable Rate for the ACSI Channel} \label{sec:Achievable ACSI}
In this section we address the case of causal state information at the transmitter, i.e., the ACSI channel model,  as previously defined Section \ref{acsi_def}.
We next show that, unlike the AGP channel model, if the set $\mathcal{D}$ includes positive delays, the encoder can ignore the side information with no loss of optimality in terms of achievable rates. We next formalize and prove the above statement.
\begin{theorem}
If $d_{max}>0$, then the capacity of the ACSI channel is given by
\begin{equation}
C=\max_{P_X} I(X;Y),
\end{equation}
where $P_{Y|X}(y|x)=\sum_s P_S(s)P_{Y|X,S}(y|x,s)$.
\end{theorem}
\begin{IEEEproof}
 This is a direct consequence of the fact the ACSI channel setup with $d_{max}>0$ is inferior capacity-wise to the {\it synchronous} setup of strictly causal side information, i.e., at time instant $i$, the encoder observes $(s_1,\ldots,s_{i-1})$ for which it was shown in \cite{maric} that the side information can be ignored without loss of optimality.
\end{IEEEproof}

We note that if $d_{max}=0$, then the synchronous counterpart of this setup for $d<0$ is that of a limited lookahead analyzed in \cite[Section~VI, Theorem~8]{WeissmanElGamal} which results in a multi-letter expression for the capacity.
The encoding scheme we use for the case $d_{max}=0$ is similar to that of the AGP channel described in Section \ref{sec:Achievable AGP}, in the sense that the transmitter splits the timeline $\{1,\ldots,n\}$ into $D=d_{min}+1$ segments. In each of the segments, the coding scheme that corresponds to the appropriate lookahead $d$ \cite{WeissmanElGamal} is applied (and in the segment corresponding to $d=0$ Shannon's causal scheme \cite{shannon} is applied). The resulting achievable rate does not have a single-letter expression and is omitted for the sake of brevity.

\section{Asynchronous Channels with States Non-Causally Available to the Transmitter and the Receiver}\label{sec:Achievable ACSITR}
In this section, we derive the capacity formula of asynchronous channels with states non-causally available both at the transmitter and the receiver (see Fig.\ \ref{Delay_trans_receiv_fig} and Section \ref{ACSITR_def}).  We note that since the decoder knows the side information (and can deduce the actual delay), both the encoder and the decoder rely on the side information sequence in their enccoding/decoding strategies. This is the fundamental difference from the AGP channel model, in which the decoder can only rely on the statistics of the state-sequence if the coding scheme does not include sending the side information to the decoder.
Finally, we show that a coding scheme that considers all possible side information symbols for all possible delays is capacity achieving.

\begin{theorem}\label{capacity_ACSITR}
Let $\mathcal{D}$ be a set of possible delays and $D=|\mathcal{D}|$.
The capacity of the asynchronous channel with states non-causally available at the transmitter and the receiver and a channel conditional distribution $P_{Y|X,S}$ is
\begin{flalign}\label{capacity_ACSITR_eq}
C_{ACSITR} = \max_{P(x|v)}\min _{d\in \mathcal{D}} I_d(X;Y|V)
\end{flalign}
where $V\in \mathcal {S}^{D}$ is a random variable distributed according to $P_V(v)=\prod_{i=1}^{D} P_S(v_i)$, and
\begin{flalign}\label{acsitr_prob_def}
&P_d(x,y|v) = P(x|v)P(y|x,v_{d_{max}-d+1})\nonumber\\
&P_d(y|v)=\sum_{x\in\mathcal{X}}P_d(x,y|v)
\end{flalign}
where $v_{d_{max}-d+1}$ is the $(d_{max}-d+1)^{th}$ entry in the vector $v$.
Additionally,
\begin{flalign}
I_d(X;Y|V)=\sum_{x\in\mathcal{X},y\in\mathcal{Y},v\in\mathcal{V}}P(v)P_d(x,y|v)\log\left(\frac{P_d(x,y|v)}{P(x|v)P_d(y|v)}\right).
\end{flalign}
\end{theorem}
The achievability coding scheme consists of a "strategy letters" coding scheme \cite{Salehi1992}. It is implemented by using the sequence $V^n$ as the state-sequence which the strategy maps. The detailed proof which consists of the achievability part and the converse part is included in Appendix \ref{ACSITR_Capacity_Proof}. Further, we note that a naive rate-splitting coding scheme which uses the sequence $V^n$ as a time-sharing sequence may lead to suboptimal results since each sub-message is separately reconstructed under all possible delays. This degradation follows from the independence between sub-messages in rate-splitting coding scheme which prohibits us from using redundancy in one sub-message to help us decoding another sub-message.

We remark that in contrast to the synchronous models, in which causal and non-causal knowledge of the state-sequence at both the encoder and decoder yield the same channel capacity, in the asynchronous setup the capacities of these models do not necessarily coincide.

In addition, one can consider a different setup in which the delay $d$ symbolizes the presence of a jitter. The jitter is modeled by a delay that randomly changes every sub-block of a sufficiently large size that allows the decoder to find the delay in the sub-block with an error probability that decays with the block length. It can be shown that in this setup, if the delays are i.i.d. random variables distributed over the set $\mathcal{D}$, the minimization over the delay $d$ in (\ref{capacity_ACSITR_eq}) can be replaced with an expectation over the delay $d$.

Finally, the result generalizes straightforwardly to state dependent compound channels with state information at the transmitter and receiver. Specifically, let $\Theta$ be a finite set of channels from $\mathcal{X}\times\mathcal{S}$ to $\mathcal{Y}$, and let $P_{\theta}(y|x,s)$ denote the transition probability of channel $\theta$, as before $X,S,Y$ denote the channel input, channel state, and channel output, respectively.
\begin{corollary}
The capacity of the state dependent compound channel is given by
\begin{flalign}
C=\max_{P(x|s)}\min_{\theta\in \Theta} I_{\theta}(X;Y|S)
\end{flalign}
where
\[P_{\theta}(x,y|s)=P(x|s)P_{\theta}(y|x,s).\]
\end{corollary}

\section{Conclusion}\label{sec:Conclusion}
In this paper we presented several asynchronous channel models that include side information at the transmitter and/or receiver. We derived an achievable rate for the AGP channel using an encoding scheme which combines binning and time sharing. We then generalized this lower bound by representing the AGP channel as a compound channel. Further, we proved that although the side information is known asynchronously, it is still of value and can be exploited. We further discussed the ACSI channel in which the side information is available asynchronously and causally at the transmitter. We proved that if the delay can take positive values then the side information does not increase the capacity of the ACSI channel.
Finally, we established a single-letter expression for the capacity of asynchronous channels with side information at both the transmitter and receiver.

\begin{appendices}
\section{}\label{AGP_A1}
In this section, we present the coding scheme of Theorem \ref{theorem_AGP1} (AGP channel) and analyze the resulting average probability of error.
We present the proof for the case $d\in\{0,1\}$, i.e., $d_{max}=1, d_{min}=0$, which can be easily generalized for any finite $d_{min},d_{max}$, and for simplicity, we assume that the alphabets are finite.
\newline


\textbf{Codebook Generation:}
Fix $P_{U|A}$ and  $P_{X|U,A}$, and let $P_U(u)=\sum_{a\in\mathcal{S}}P_{U|A}(u|a)P_S(a)$. For each message $m\in\left\{1,\ldots,2^{nR}\right\}$ generate a subcodebook (a bin) consisting of $2^{nJ}$ codewords of length $n$, $u^n(m,k)$, $k\in\{1,\ldots,2^{nJ}\}$ according to $\prod_{i=1}^n P_U(u_i)$. We denote the subcodebook of message $m$ by $C(m)$, that is, $C(m)=\{u^n(m,k)\}_{k=1}^{2^{nJ}}$.


\textbf{Encoding: }
Upon observing the sequence of states $a^n$ (which is a possibly delayed version of $s^n$), to send message $m$ choose $u^n\in C(m)$ whose first $\frac{n}{2}$ symbols are jointly typical with $a_1^{\frac{n}{2}}$ and whose subsequent $\frac{n}{2}-1$ symbols are jointly typical with $a_{\frac{n}{2}+2}^n$.
The encoder then generates  $x_1^{\frac{n}{2}}$ i.i.d. given $(u_1^{\frac{n}{2}},a_1^{\frac{n}{2}})$, that is, according to $\prod_{i=1}^{\frac{n}{2}}P_{X|U,A}(x_i|u_i,a_i)$. The next $\frac{n}{2}-1$ symbols, $x_{\frac{n}{2}+1}^{n-1}$, are generated i.i.d. given $(u_{n/2+1}^n,a_{n/2+1}^n)$, that is, according to $\prod_{i=\frac{n}{2}+1}^{n-1}P_{X|U,A}(x_i|u_i,a_{i+1})$. The last symbol $x_n$ is chosen arbitrarily.


\textbf{Decoding:} If $d=0$, find $\ell$ such that there exists $u^n\in C(\ell)$  that satisfies
\begin{flalign}
	&\left( u_1^{\frac{n}{2}} , y_1^{\frac{n}{2}} \right)\in T_{p_1,\epsilon}^{\frac{n}{2}}(U,Y) \nonumber \\
	&\qquad\qquad\qquad \text{ and }\left( u_{\frac{n}{2}+1}^{n-1}, y_{\frac{n}{2}+1}^{n-1} \right)\in T_{p_2,\epsilon}^{\frac{n}{2}-1}(U,Y).
\end{flalign}
If $d=1$, find $\ell$ such that there exists $u^n\in C(\ell)$  that satisfies
\begin{flalign}
	&\left( u_{\frac{n}{2}+1}^{n-1} , y_{\frac{n}{2}+1}^{n-1} \right)\in T_{p_1,\epsilon}^{\frac{n}{2}-1}(U,Y) \nonumber \\
	&\qquad\qquad\qquad \text{ and }\left( u_1^{\frac{n}{2}} , y_1^{\frac{n}{2}} \right)\in T_{p_2,\epsilon}^{\frac{n}{2}}(U,Y)
\end{flalign}
where, $p_1(u,y),p_2(u,y)$ are as in (\ref{eq:prob}).\newline
If such an $\ell$ does not exist, or if there is more than one such $\ell$, an error is declared.

\textbf{The analysis of the average probability of error:}
Assume without loss of generality that message $1$ is sent. An error occurs if one or more of the following events take place:
\begin{enumerate}
	\item There is no $u^n\in C(1)$ such that
		\begin{flalign} \label{eq:typ}
		&\left(u_1^{\frac{n}{2}},a_1^{\frac{n}{2}}\right) \in T^{\frac{n}{2}}_{\epsilon}(U,A)  \nonumber \\  	&\qquad\qquad\qquad \text{ and }  \left(u_{\frac{n}{2}+1}^{n-1},a_{\frac{n}{2}+2}^{n}\right) \in T^{\frac{n}{2}-1}_{\epsilon}(U,A).
\end{flalign}
	We denote this event by $\mathcal{E}_1$.
	\item Denote the vector $u^n\in C(1)$ that satisfies (\ref{eq:typ}) by $\tilde{u}^n$. Consider the event
\begin{flalign}\label{eq:event1}
	&\left( \tilde{u}_1^{\frac{n}{2}} , y_1^{\frac{n}{2}} \right)\notin T_{p_1,\epsilon}^{\frac{n}{2}}(U,Y) \nonumber \\
	&\qquad\qquad\qquad\text{ or } \left( \tilde{u}_{\frac{n}{2}+1}^{n-1} , y_{\frac{n}{2}+1}^{n-1} \right)\notin T_{p_2,\epsilon}^{\frac{n}{2}-1}(U,Y)
\end{flalign}	
given that $d=0$.
Additionally, consider the event
\begin{flalign}\label{eq:event2}
	&\left( \tilde{u}_{\frac{n}{2}+1}^{n-1}, y_{\frac{n}{2}+1}^{n-1} \right)\notin T_{p_1,\epsilon}^{\frac{n}{2}-1}(U,Y) \nonumber \\
	&\qquad\qquad\qquad\text{ or }\left( \tilde{u}_1^{\frac{n}{2}} , y_1^{\frac{n}{2}} \right)\notin T_{p_2,\epsilon}^{\frac{n}{2}}(U,Y)
\end{flalign}
given that $d=1$.

	We denote  events (\ref{eq:event1}) and (\ref{eq:event2}) by $\mathcal{E}_{2,1}$ and $\mathcal{E}_{2,2}$, respectively, and their union by $\mathcal{E}_2$.
	\item Given that $d=0$ there exists $m'\neq 1$ and $u^n\in C(m')$ such that
	\begin{flalign} \label{eq:e31}
	&\left( u_1^{\frac{n}{2}} , y_1^{\frac{n}{2}} \right)\in T_{p_1,\epsilon}^{\frac{n}{2}}(U,Y) \nonumber \\
	&\qquad\qquad\qquad\text{ and }\left( u_{\frac{n}{2}+1}^{n-1} , y_{\frac{n}{2}+1}^{n-1} \right)\in T_{p_2,\epsilon}^{\frac{n}{2}-1}(U,Y).
	\end{flalign}
Alternatively, if $d=1$ there exists $m'\neq 1$ and $u^n\in C(m')$ such that
\begin{flalign} \label{eq:e32}
	&\left( u_{\frac{n}{2}+1}^{n-1}, y_{\frac{n}{2}+1}^{n-1} \right)\in T_{p_1,\epsilon}^{\frac{n}{2}-1}(U,Y)\nonumber \\
	&\qquad\qquad\qquad\text{ and }\left( u_1^{\frac{n}{2}} , y_1^{\frac{n}{2}} \right)\in T_{p_2,\epsilon}^{\frac{n}{2}}(U,Y).
\end{flalign}
	We denote this event by $\mathcal{E}_3$.
	\end{enumerate}
	The error event is the union of $\mathcal{E}_i$, $i=1,2,3$, thus by the union bound,
\begin{flalign}
	\Pr(\mathcal{E}) = \Pr(\mathcal{E}_1  \cup \mathcal{E}_2 \cup \mathcal{E}_3) \leq \Pr(\mathcal{E}_1)+\Pr(\mathcal{E}_2)+P( \mathcal{E}_3).
\end{flalign}
By the covering lemma \cite[p. 62]{NetworkInformationTheory}, if $J>I(U;A)$ then $\Pr(\mathcal{E}_1) \rightarrow 0$ as $n \rightarrow \infty$.

By definition of $\Pr(\mathcal{E}_2)$, it follows that
\begin{flalign}
&\Pr(\mathcal{E}_2)= \Pr(\mathcal{E}_{2,1}\cup \mathcal{E}_{2,2})=\mathbbm{1}_{\{d=0\}}\cdot\Pr(\mathcal{E}_{2,1})+\mathbbm{1}_{\{d=1\}}\cdot\Pr(\mathcal{E}_{2,2}).
\end{flalign}
Now, assume without loss of generality that $d=1$, from the conditional typicality lemma \cite[p. 27]{NetworkInformationTheory} $\Pr(\mathcal{E}_2)\rightarrow 0$ as $n\rightarrow \infty $. Similarly, if $d=1$, then $\Pr(\mathcal{E}_{2,2})\rightarrow 0$ as $n\rightarrow \infty $. Therefore, $\Pr(\mathcal{E}_{2})\rightarrow 0$ as $n\rightarrow \infty $.

It remains to bound $\Pr(\mathcal{E}_3)$. Let $B_1$ be the set
 \begin{flalign}
 &\left\{\left(u^n,y^n\right): \left( u_1^{\frac{n}{2}} , y_1^{\frac{n}{2}} \right)\in T_{p_1,\epsilon}^{\frac{n}{2}}(U,Y) \right.\nonumber\\
 &\qquad\qquad\qquad\qquad\text{ and }\left.\left( u_{\frac{n}{2}+1}^{n-1} , y_{\frac{n}{2}+1}^{n-1} \right)\in T_{p_2,\epsilon}^{\frac{n}{2}-1}(U,Y)\right\},
 \end{flalign}
 and let $B_2$ be the set
 \begin{flalign}
 &\left\{\left(u^n,y^n\right): \left( u_1^{\frac{n}{2}} , y_1^{\frac{n}{2}} \right)\in T_{p_2,\epsilon}^{\frac{n}{2}}(U,Y) \right.\nonumber\\
 &\qquad\qquad\qquad\qquad\text{ and }\left.\left( u_{\frac{n}{2}+1}^{n-1} , y_{\frac{n}{2}+1}^{n-1} \right)\in T_{p_1,\epsilon}^{\frac{n}{2}-1}(U,Y)\right\}.
 \end{flalign}
 In addition, define $p_1(y) = \sum_{u}p_1(u,y)$, and $p_2(y) = \sum_{u}p_2(u,y)$, where $p_1(u,y),p_2(u,y)$ are as in (\ref{eq:prob}).
Suppose that $U^n$ is generated i.i.d. according to $P_U(u)$, additionally, suppose that $y_1^{\frac{n}{2}}\in T_{\epsilon,p_1}^{\frac{n}{2}}(Y)$ and $y_{\frac{n}{2}+1}^{n-1}\in T_{\epsilon,p_2}^{\frac{n}{2}-1}(Y)$, then
\begin{flalign}\label{ind1}
  \Pr\{(U^n,y^n)\in B_1\}&\leq\sum_{u^n:(u^n,y^n)\in B_1} P(u^n)\nonumber\\
  &\leq\sum_{u^n:(u^n,y^n)\in B_1}P(u^{n-1})\nonumber\\
  &\stackrel{(a)}\leq\sum_{u^n:(u^n,y^n)\in B_1} 2^{-(n-1)\left[H(U)-\delta(\epsilon)\right]}\nonumber\\
&\stackrel{(b)}\leq 2^{\frac{n}{2}H_{p_1}(U|Y)+\left(\frac{n}{2}-1\right)H_{p_2}(U|Y)+(n-1)\delta'(\epsilon)}  2^{-(n-1)\left[H(U)-\delta(\epsilon)\right]}\nonumber\\
&\leq 2^{-\left[\left(\frac{n}{2}-1 \right)I_{p_1}(U;Y)+\left(\frac{n}{2}-1 \right)I_{p_2}(U;Y)-(n-1)\delta''(\epsilon)\right]}
\end{flalign}
where $(a)$ follows since $U^n$ is generated i.i.d., and (b) follows from the definition of $B_1$, the fact that  $u_1^{\frac{n}{2}}$ and $u_{\frac{n}{2}+1}^n$ are statistically independent, and from Theorem 1.3 in \cite{Kramer}. In addition, $\delta(\epsilon),\delta'(\epsilon),\delta''(\epsilon)$ are functions of $\epsilon$ which vanish as $\epsilon$ tends to $0$.

Similarly, if $y_1^{\frac{n}{2}}\in T_{\epsilon,p_2}^{\frac{n}{2}}(Y)$ and $y_{\frac{n}{2}+1}^{n-1}\in T_{\epsilon,p_1}^{\frac{n}{2}-1}(Y)$, and $U^n$ is generated i.i.d. according to $P_U(u)$, then
\begin{flalign}\label{ind2}
  &\Pr\left((U^n,y^n)\in B_2\right)
  \leq 2^{-\left[\left(\frac{n}{2}-1 \right)I_{p_1}(U;Y)+\left(\frac{n}{2}-1 \right)I_{p_2}(U;Y)-(n-1)\delta''(\epsilon)\right]}.
\end{flalign}

Now, let $u_{i,j}^n$ be the $j$th codeword in $C(i)$, using (\ref{ind1}) and (\ref{ind2}), we get
\begin{flalign}
&\Pr(\mathcal{E}_3)\leq \mathbbm{1}_{\{d=0\}}\cdot  \sum_{i=2}^{2^{nR}}\sum_{j=1}^{2^{nJ}} P\left((u_{i,j}^n,y^n)\in B_1\right)
+\mathbbm{1}_{\{d=1\}}\cdot \sum_{i=2}^{2^{nR}}\sum_{j=1}^{2^{nJ}} P\left((u_{i,j}^n,y^n)\in B_2\right)  \nonumber\\
&\leq  \mathbbm{1}_{\{d=0\}}\cdot \sum_{i=2}^{2^{nR}}\sum_{j=1}^{2^{nJ}} P\left((u_{i,j}^n,y^n)\in B_1 \right)+\mathbbm{1}_{\{d=1\}}\cdot \sum_{i=2}^{2^{nR}}\sum_{j=1}^{2^{nJ}} P\left((u_{i,j}^n,y^n)\in B_2 \right)  \nonumber\\
&\leq 2^{n(R+J)}2^{-\left[\left(\frac{n}{2}-1 \right)I_{p_1}(U;Y)+\cdot \left(\frac{n}{2}-1 \right)I_{p_2}(U;Y)-(n-1)\delta''(\epsilon)-1\right]}.
\end{flalign}
Hence, as $R+J<\frac{1}{2}I_{p_1}(U;Y)+\frac{1}{2}I_{p_2}(U;Y)$ and $J>I(U;A)$ $\Pr(\mathcal{E}) \rightarrow 0$ as $n \rightarrow \infty$.

\section{}\label{AGP_A2}
In this section, we present the coding scheme of the AGP channel which corresponds to Theorem 2 and analyze the probability of error of this coding scheme.

For the sake of clarity, we present the proof for the case $d_{min}=0,d_{max}=1$, i.e., we prove that the rate in (\ref{AGP_lower_rate}) is achievable in this case. We note that both the encoder and the decoder know the time sharing sequence $t^{n}\in T_{\epsilon}^n(T)$. Additionally, as mentioned before, we can assume that the decoder knows the actual delay $d$ in the channel.  Finally, we ignore the end effects in our notations, since the first/last symbols do not affect the asymptotic performance in terms of the achievable reliable rates.

Now, since the decoder can find the delay $d$, and the encoder does not know the delay,  we must be able to decode the message for both delays $d=0,1$ simultaneously. Therefore, we can represent the AGP channel as a multicast channel as appears in Fig.\ \ref{AGP_multicast_fig}.
Let channel $1$  be the channel with $d=0$ and channel $2$ be the channel where the state is delayed that is $d=1$, and let
$y_1^n$ and $y_2^n$ be the outputs of channels $1$ and $2$ respectively.

Denote the statistically dependent $D-$tuples: $V_1^n=(S_{i-d_{min}},\ldots,S_{i+d_{max}})_{i=1}^n$, where
\begin{flalign}\label{Pv}
P_V(v)=\prod_{i=1}^DP_{S}(v_i).
\end{flalign}

\textbf{Codebook Generation:}
Set $P_T$ and $P_{W,U_1,U_2,X|T,V}$, where $\mathcal{V}=\mathcal{S}^D$.
\begin{itemize}
    \item Generate a sequence $t^n$ according to $\prod_{i=1}^nP_T(t_i)$.
	\item For each message $m\in\{1,\ldots,2^{nR}\}$, generate $2^{nT_0}$ codewords, $\{w^{n}(m,l_0)\}_{l_0=1}^{2^{nT_0}}$, according to
	$\prod_{i=1}^{n} P_{W|T}(w_i|t_i)$.
	\item For each  sequence $w^n(m,l_0)$, generate $2^{nT_1}$ codewords, $\{u_1^{n}(m,l_0,l_1)\}_{l_1=1}^{2^{nT_1}}$, according to
	$\prod_{i=1}^{n} P_{U_1|W,T}(u_{1,i}|w_i,t_i)$.
	\item For each sequence $w^n(m,l_0)$, generate $2^{nT_2}$ codewords, $\{u_2^{n}(m,l_0,l_2)\}_{l_2=1}^{2^{nT_2}}$, according to
	 $\prod_{i=1}^{n} P_{U_2|W,T}(u_{2,i}|w_i,t_i)$.
\end{itemize}

\textbf{Encoding:}
Let $a^n$ be the possibly delayed state sequence, generate the sequence $v^n$ from the sequence $a^n$ in the following manner
\begin{flalign}
v_i=(a_{i-d_{min}},\ldots,a_{i+d_{max}})=(a_{i},a_{i+1})
\end{flalign}
where the last equality is follows since $d_{min}=0$ and $d_{max}=1$,  $a_{n+1}$ is chosen arbitrarily.

To send message $m$,
\begin{itemize}
	\item Find $l_0\in \{1,\ldots,2^{nT_0}\}$ such that
\begin{flalign}
\left(w^{n}(m,l_0),v^{n},t^n\right)\in T_{\epsilon}^{n}(W,V,T).
\end{flalign}
\item Next, find $l_1\in\{1,\ldots,2^{nT_1}\}$ and $l_2\in\{1,\ldots,2^{nT_2}\}$ such that
\begin{flalign}
&\left(w^{n}(m,l_0),u_1^{n}(m,l_0,l_1),u_2^{n}(m,l_0,l_2),v^{n},t^n\right)\in T_{\epsilon}^{n}(W,U_1,U_2,V,T).
\end{flalign}
\item Generate $X^{n}$ according to,
 \begin{flalign}
 &\prod_{i=1}^{n} P_{X|W,U_1,U_2,V,T}
 \left(x_i|w_i(m,l_0),u_{1,i}(m,l_0,l_1),u_{2,i}(m,l_0,l_2),v_i,t_i\right),
 \end{flalign}
 and transmit $X^n$.
\end{itemize}

\textbf{Decoding:}
To decode the message,
\begin{itemize}
	\item Decoder 1: look for $\hat{m}_1\in\{1,\ldots,2^{nR}\}$, $\hat{l}_0\in\{1,\ldots,2^{nT_0}\}$ and $\hat{l}_1\in\{1,\ldots,2^{nT_1}\}$ such that,
\begin{flalign}
&\left(w^{n}(\hat{m}_1,\hat{l}_0),u_1^{n}(\hat{m}_1,\hat{l}_0,\hat{l}_1),y_1^{n},t^n\right)\in T_{\epsilon}^{n}(W,U_1,Y_1,T).
\end{flalign}
If there is only one such $\hat{m}_1$, it is the decoded message, otherwise an error is declared.
	\item Decoder 2: look for $\hat{m}_2\in\{1,\ldots,2^{nR}\}$, $\hat{l}_0\in\{1,\ldots,2^{nT_0}\}$ and $\hat{l}_2\in\{1,\ldots,2^{nT_2}\}$ such that,
\begin{flalign}
&\left(w^{n}(\hat{m}_2,\hat{l}_0),u_2^{n}(\hat{m}_2,\hat{l}_0,\hat{l}_2),y_2^{n},t^n\right)\in T_{\epsilon}^{n}(W,U_2,Y_2,T).
\end{flalign}
If there is only one such $\hat{m}_2$, it is the decoded message, otherwise an error is declared.
\end{itemize}

\textbf{The analysis of the average probability of error:}
Suppose without loss of generality that the message $m=1$ was sent.
An error is made if one of the following events occurs,
\begin{enumerate}

\item The sequence $t^{n}$ is not in $ T_{\epsilon}^{n} (T)$, we denote this event by $\mathcal{E}_{t}$.

\item The sequence $v^{n}$ is not in $ T_{\epsilon}^{n} (V)$, we denote this event by $\mathcal{E}_{v}$.

	\item There is no $l_0 \in \{1,\ldots,2^{nT_0}\}$ such that,
	\begin{flalign}\label{e1}
\left(w^{n}(1,l_0),v^{n},t^n \right)\in T_{\epsilon}^{n}(W,V,T).
\end{flalign}
We denote this event by $\mathcal{E}_{e,1}$.

	\item There are no $l_1\in\{1,\ldots,2^{nT_1}\}$ and $l_2\in\{1,\ldots,2^{nT_2}\}$ such that,
\begin{flalign}
&\left(w^{n}(1,l_0),u_1^{n}(1,l_0,l_1),u_2^{n}(1,l_0,l_2),v^{n},t^n\right)
\in T_{\epsilon}^{n}(W,U_1,U_2,V,T).
\end{flalign}
We denote this event by $\mathcal{E}_{e,2}$.

\item The indices $l_0,l_1$ were chosen, but
\begin{flalign}
&\left(w^{n}(1,l_0),u_1^{n}(1,l_0,l_1),y_1^{n},t^n\right)\notin T_{\epsilon}^{n}(W,U_1,Y_1,T).
\end{flalign}
We denote this event by $\mathcal{E}_{d_1,1}$.

	\item There exists $m'\neq 1$,
\begin{flalign}
&\left(w^{n}(m',l'_0),u_1^{n}(m',l'_0,l'_1),y_1^{n},t^n\right)\in T_{\epsilon}^{n}(W,U_1,Y_1,T).
\end{flalign}
for some $l'_0\in\{1,\ldots,2^{nT_0}\}$ and $l'_1\in\{1,\ldots,2^{nT_1}\}$. We denote this event by
$\mathcal{E}_{d_1,2}$.

\item The indices $l_0,l_2$ were chosen, but
\begin{flalign}
&\left(w^{n}(1,l_0),u_2^{n}(1,l_0,l_1),y_2^{n},t^n\right)\notin T_{\epsilon}^{n}(W,U_2,Y_2,T).
\end{flalign}
We denote this event by $\mathcal{E}_{d_2,1}$.

	\item There exists $m'\neq 1$ such that,
\begin{flalign}
&\left(w^{n}(m',l'_0),u_2^{n}(m',l'_0,l'_1),y_2^{n},t^n\right)\in T_{\epsilon}^{n}(W,U_2,Y_2,T).
\end{flalign}
for some $l'_0\in\{1,\ldots,2^{nT_0}\}$ and $l'_2\in\{1,\ldots,2^{nT_2}\}$. We denote this event by
$\mathcal{E}_{d_2,2}$.
\end{enumerate}
Let,
\begin{flalign}
&\mathcal{E}_{tv}=\mathcal{E}_{t}\cup \mathcal{E}_{v}\nonumber\\
&\mathcal{E}_{e}=\mathcal{E}_{e,1}\cup \mathcal{E}_{e,2}\nonumber\\
&\mathcal{E}_{d_1}=\mathcal{E}_{d_1,1}\cup \mathcal{E}_{d_1,2}\nonumber\\
&\mathcal{E}_{d_2}=\mathcal{E}_{d_2,1}\cup \mathcal{E}_{d_2,2}.
\end{flalign}
There average error probability is further bounded by,
\begin{flalign}
\Pr(\mathcal{E})&\leq \Pr( \mathcal{E}_{t}\cup \mathcal{E}_{v}\cup \mathcal{E}_{e}\cup \mathcal{E}_{d_1}\cup \mathcal{E}_{d_2})\nonumber\\
&\leq \Pr(\mathcal{E}_{t})+ \Pr(\mathcal{E}_{v})+ \Pr( \mathcal{E}_{tv}^c\cap \mathcal{E}_{e,1})\nonumber\\
&\quad+ \Pr( \mathcal{E}_{tv}^c\cap \mathcal{E}_{e,1}^c\cap \mathcal{E}_{e,2})+\Pr(\mathcal{E}_{tv}^c\cap \mathcal{E}_{e}^c \cap \mathcal{E}_{d_1,1})\nonumber\\
&\quad+\Pr(\mathcal{E}_{d_1,2})+\Pr(\mathcal{E}_{tv}^c\cap \mathcal{E}_{e}^c \cap \mathcal{E}_{d_2,1})+\Pr(\mathcal{E}_{d_2,2}).
\end{flalign}

By the LLN, $\Pr(\mathcal{E}_{t})\rightarrow 0$ as $n\rightarrow\infty$. In addition, from the stationarity and ergodicity of $v^{n}$ we infer that $\Pr(\mathcal{E}_{v})\rightarrow 0$ as $n\rightarrow\infty$.

By the covering lemma\footnote{Note that this lemma does not demand $V^n$ to be statistically independent, the only assumption is that $V^n$ is a typical sequence with respect to $P_V$. This is also true for the other lemmas and Theorems which we use in this proof.} \cite[p. 62]{NetworkInformationTheory}, $\Pr(\mathcal{E}_{tv}^c\cap \mathcal{E}_{e,1})\rightarrow 0$ as $n\rightarrow 0$, if
\begin{flalign}\label{ineq0}
T_0 \geq I(W;V|T).
\end{flalign}

An immediate extension of \cite[Appendix A]{Nair}, yields that if
\begin{flalign}\label{ineq1}
&T_1 > I(U_1;V|W,T)\nonumber\\
&T_2 > I(U_2;V|W,T)\nonumber\\
&T_1+T_2 > I(U_1;V|W,T)+I(U_2;V|W,T)+I(U_1;U_2|V,W,T)
\end{flalign}
then $\Pr(\mathcal{E}_{tv}^c\cap \mathcal{E}_{e,1}^c\cap \mathcal{E}_{e,2})\rightarrow 0$ as $n\rightarrow\infty$.

Next, an immediate extension of the Markov Lemma \cite[Lemma 15.8.1]{CT}, gives that $\Pr(\mathcal{E}_{tv}^c\cap \mathcal{E}_{e}^c \cap \mathcal{E}_{d_1,1})\rightarrow 0$ and $\Pr(\mathcal{E}_{tv}^c\cap \mathcal{E}_{e}^c \cap \mathcal{E}_{d_2,1})\rightarrow 0$ as
$n\rightarrow\infty$.

Finally, by \cite[Theorem 7.1]{Kramer}, if
\begin{flalign}\label{ineq2}
&R+T_0+T_1\leq I(W,U_1;Y_1|T)\nonumber\\
&R+T_0+T_2\leq I(W,U_2;Y_2|T)
\end{flalign}
then $\Pr(\mathcal{E}_{d_1,2})\rightarrow 0$ and $\Pr(\mathcal{E}_{d_2,2})\rightarrow 0$ as $n\rightarrow\infty$.

Performing  Fourier-Motzkin Elimination on (\ref{ineq0})-(\ref{ineq2}) yields the rate  (\ref{AGP_lower_rate}).

\section{}\label{ACSITR_Capacity_Proof}
In this section, we present the coding scheme of Theorem \ref{capacity_ACSITR} and analyze the respective probability of error. The converse part of Theorem \ref{capacity_ACSITR} is presented as well.

\begin{IEEEproof}[Proof of the Achievability Part of Theorem \ref{capacity_ACSITR}]
As before,  the decoder can deduce $d$ from $y^n$ with arbitrarily low probability of error, so we assume that $d$ is known at the decoder.

Suppose that $|\mathcal{S}|<\infty$. Let $v_i=(s_{i-d_{max}},\ldots,s_{i+d_{min}})$, where $s_{i-d}$ are arbitrary for all $i\in \{1,\ldots,n\}$ such that $i-d\notin\{1,\ldots,n\}$. Additionally, denote $\mathcal{V}=\mathcal{S}^D$ where $D=|\mathcal{D}|$. Note that since both the encoder and decoder know the sequence $s^n$ and the set of possible delays, each can build the sequence $v^n$.

\textbf{Codebook Generation:} Fix a conditional p.m.f. $P(x|v)$. Further, order the symbols of the alphabet $\mathcal{V}$ in some manner and let $N: \mathcal{V}\rightarrow \{1,\ldots,|\mathcal{V}|\}$ be the chosen ordering function.
For each $m\in\mathcal{M}$ denote by $b(m)$ a matrix of dimensions $|\mathcal{V}|\times n$ which is generated in the following manner. Let $b(m,i)$ be the $i$-th column of $b(m)$, and let $b(m,i,N(v))$ be the $N(v)$-th entry in the column vector $b(m,i)$.
Each $b(m)$ is generated according an i.i.d. distribution, that is,
\begin{flalign}
P_B(b)=\prod_{i=1}^nP(b_i),
\end{flalign}
where $b_i$ is the $i$-th column of the matrix  $b$.
Denote by $b_{i,N(v)}$ the $N(v)$-th entry of the column vector $b_i$. Each of the column vectors $b_i$ is generated according to
\begin{flalign}
P(b_i)=\prod_{v\in\mathcal{V}}P(b_{i,N(v)})
\end{flalign}
where
\begin{flalign}
P(b_{i,N(v)})=P_{X|V}(b_{i,N(v)}|v).
\end{flalign}

\textbf{Encoding:} To send  $m\in \mathcal{M}$, in each time instant $i$, the transmitter sends $x_i=b(m,i,N(v_i))$.

\textbf{Decoding:}
Let $x_i(m,v^n)=b(m,i,N(v_i))$ and let $x^n(m,v^n)=(x_i(m,v^n))_{i=1}^n$.
Upon receiving $y^n$ the decoder looks for $\hat{m}\in\mathcal{M}$ such that
\begin{flalign}
(x^n(m,v^n),y^n)\in T_{d,\epsilon}^n (X,Y|v^n)
\end{flalign}
where $P_d(x,y|v)$ is defined in (\ref{acsitr_prob_def}).

\textbf{Analysis of Probability of Error:}
Suppose the message $m=1$ was sent.
An error is made if one of the following events occurs:
\begin{flalign}
&\mathcal{E}_1=\{v^n\notin T_{\epsilon}^n(V)\}\nonumber\\
&\mathcal{E}_2=\{ x^n(1,v^n)\notin T_{\epsilon}^n(X|V=v)\}\nonumber\\
&\mathcal{E}_3=\{(x^{n}(1,v^n),y^{n})\notin T_{d,\epsilon}^{n}(X,Y|v^n)\}\nonumber\\
&\mathcal{E}_4=\begin{Bmatrix}
\exists  \tilde{m}\neq 1 \text{ s.t. } \\
 (x^{n}(\tilde{m},v^n),y^{n})\in T_{d,\epsilon}^{n}(X,Y|v^n)
\end{Bmatrix}.
\end{flalign}
Therefore, the average probability of error $\Pr(\mathcal{E})$  satisfies
\begin{flalign}
\Pr(\mathcal{E})&=\Pr(\mathcal{\mathcal{E}}_1\cup \mathcal{E}_2\cup \mathcal{E}_3\cup \mathcal{E}_4 )\leq \Pr(\mathcal{E}_1)+\Pr(\mathcal{E}_1^c \cap \mathcal{E}_2)+\Pr(\mathcal{E}_1^c \cap \mathcal{E}_2^c \cap \mathcal{E}_3)+\Pr(\mathcal{E}_4).
\end{flalign}
By the ergodicity and stationarity of $v^n$, $\Pr(\mathcal{E}_1)\rightarrow0$ as $n\rightarrow\infty$. By the conditional typicality lemma \cite[p. 27]{NetworkInformationTheory}, $\Pr(\mathcal{E}_1^c \cap \mathcal{E}_2)\rightarrow0$ and
$\Pr(\mathcal{E}_1^c \cap \mathcal{E}_2^c \cap \mathcal{E}_3)\rightarrow0$  as $n\rightarrow\infty$.

We next prove that the sequence $y^n$ is memoryless given the sequence $v^n$ and the delay $d$ (which is assumed to be known at the decoder).
\begin{flalign}
P_d(y^n|v^n)&= \sum_{x^n\in \mathcal{X}^n} P_d(y^n,x^n|v^n)\nonumber\\
&=\sum_{x^n\in \mathcal{X}^n} P_d(y^n|x^n,v^n)P(x^n|v^n)\nonumber\\
&=\sum_{x^n\in \mathcal{X}^n} \prod_{i=1}^n P_d(y_i|x_i,v_i)p(x_i|v_i)\nonumber\\
&= \prod_{i=1}^n\sum_{x_i\in \mathcal{X}} P_d(y_i|x_i,v_i)p(x_i|v_i)\nonumber\\
&=\prod_{i=1}^n P_d(y_i|v_i).
\end{flalign}
Now, by the packing lemma\footnote{Note that this lemma does not demand $V^n$ to be statistically independent, the only assumption is that $V^n$ is a typical sequence with respect to $P_V$.}  \cite[p. 46]{NetworkInformationTheory}  $\Pr(\mathcal{E}_4)\rightarrow 0$ as $n\rightarrow\infty$ if
\begin{flalign}\label{ACISTR_R}
R< I_d(X;Y|V).
\end{flalign}
Additionally, since the delay is chosen in an arbitrary manner, (\ref{ACISTR_R}) must hold for every delay $d\in\mathcal{D}$.

To conclude, since $\epsilon$ is arbitrarily small, we have shown that a rate arbitrarily close to
\begin{flalign}
R=\max_{P(x|v)}\min_{d\in\mathcal{D}} I_d(X;Y|V).
\end{flalign}
is achievable.
\end{IEEEproof}

\begin{IEEEproof}[Proof of the Converse Part of Theorem \ref{capacity_ACSITR}]
Let $V_i=(S_{i-d_{max}},\ldots,S_{i+d_{min}})$ and $V_{i,j}=S_{i-d_{max}+j-1}$.
Additionally let,
\begin{flalign}
P_d(m,v^n,x^n,y^n)=P(m)P(v^n)P(x^n|v^n,m)P_d(y^n|x^n,v^n),
\end{flalign}
where
\begin{flalign}
&P(m)=2^{-nR},\newline\\
&P(v^n)=P(v_1)\cdot\prod_{i=2}^nP(v_i|v_{i-1})=\prod_{i=1}^{D}P_S(v_{1,i})\cdot\prod_{i=2}^nP_S(v_{i,D}),\newline\\
&P_d(y^n|x^n,v^n)=\prod_{i=1}^nP(y_i|x_i,v_{i,d_{max}-d+1}).
\end{flalign}
We denote information theoretic functionals of $P_d(m,v^n,x^n,y^n)$ by the subscript $d$, e.g., $H_d(M|Y^n,V^n)$.

For every sequence of $(2^{nR},n)$-codes with probability of error $P_e^{(n)}$ that vanishes as $n\rightarrow\infty$ for every $d\in \mathcal{D}$, we obtain from Fano's Inequality
\begin{flalign}
nR &= H(M) = H(M|V^n) \nonumber\\
&= H(M|V^n) - H_d(M|Y^n,V^n)+H_d(M|Y^n,V^n) \nonumber\\
&\leq  I_d(M;Y^n|V^n)+n\delta_n
\end{flalign}
where $\delta_n\rightarrow 0$ as $n\rightarrow \infty$.

We next bound the term $I_d(M;Y^n|V^n)$.
\begin{flalign}
 &I_d(M;Y^n|V^n) = \sum_{i=1}^n  I_d(M;Y_i|Y^{i-1},V^n) \nonumber\\
& = \sum_{i=1}^n  H_d(Y_i|Y^{i-1},V^n) - \sum_{i=1}^n  H_d(Y_i|M,Y^{i-1},V^n) \nonumber\\
&\stackrel{(a)}{\leq} \sum_{i=1}^n  H_d(Y_i|V_i) - \sum_{i=1}^n  H_d(Y_i|M,Y^{i-1},V^n)\nonumber\\
&\stackrel{(b)}{\leq} \sum_{i=1}^n  H_d(Y_i|V_i) - \sum_{i=1}^n  H_d(Y_i|M,X_i,Y^{i-1},V^n) \nonumber\\
&\stackrel{(c)}{=} \sum_{i=1}^n  H_d(Y_i|V_i) - \sum_{i=1}^n  H_d(Y_i|X_i,V_i) \nonumber\\
&=\sum_{i=1}^n  I_d(X_i;Y_i|V_i)
\end{flalign}
where (a) and (b) follow since conditioning reduces entropy, and (c) follows since $(M,Y^{i-1},V^{i-1},V^{n}_{i+1})-(X_i,V_i,d)-Y_i$ is a Markov chain for any \textsl{given} $d$ and all $i$. This is true since the channel is memoryless, the decoder can know the delay in the channel, and the state at time $i$, $s_{i-d}$, is included in the vector $v_i$, more specifically, $s_{i-d}=v_{i,d_{max}-d+1}$.

Let $T$  be a time sharing random variable which is distributed uniformly over $\{1,\ldots,n\}$ and independent of $V^n,X^n$ and $Y^n$, and let $X=X_T, V=V_T$ and $Y=Y_T$. Then,
\begin{flalign}
R&\leq \frac{1}{n} \sum_{i=1}^n  I_d(X_i;Y_i|V_i)+\delta_n\nonumber\\
&= I_d(X;Y|V,T)+\delta_n \nonumber\\
&=H_d(Y|V,T)-H_d(Y|X,V,T)+\delta_n\nonumber\\
&\stackrel{(a)}{=}H_d(Y|V,T)-H_d(Y|X,V)+\delta_n\nonumber\\
&\stackrel{(b)}{\leq} H_d(Y|V)-H_d(Y|X,V)+\delta_n\nonumber\\
&=I_d(X;Y|V)+\delta_n\nonumber\\
\end{flalign}
where (a) follows since $P_d(y|x,v,t)=P_d(y|x,v)=P(y|x,v_{d_{max}-d+1})$ due to the stationary nature of the channel, and (b) follows since conditioning reduces entropy.

Now, by taking the limit as $n\rightarrow\infty$ we have that,
\begin{flalign}\label{ineq_asitr}
R \leq I_d(X;Y|V).
\end{flalign}

The inequality (\ref{ineq_asitr}) holds for every $d\in\mathcal{D}$. Additionally, the encoder does not know the delay $d$ in advance, therefore
$X$ cannot depend on the delay $d$. Consequently,
\begin{flalign}
R&\leq \min_{d\in\mathcal{D}}I_d(X;Y|V)\leq \max_{P(x|v)}\min_{d\in\mathcal{D}}I_d(X;Y|V)
\end{flalign}
and this concludes the proof of Theorem \ref{capacity_ACSITR}.
\end{IEEEproof}
\end{appendices}
}

\section*{Acknowledgement}
This work was partially supported by Israel Science Foundation (ISF) grant  2013/919.
The authors would like to thank the anonymous reviewers of the Transactions on Information Theory for their helpful and constructive  comments which helped improve the content of this paper.



\begin{thebibliography}{10}
\providecommand{\url}[1]{#1}
\csname url@samestyle\endcsname
\providecommand{\newblock}{\relax}
\providecommand{\bibinfo}[2]{#2}
\providecommand{\BIBentrySTDinterwordspacing}{\spaceskip=0pt\relax}
\providecommand{\BIBentryALTinterwordstretchfactor}{4}
\providecommand{\BIBentryALTinterwordspacing}{\spaceskip=\fontdimen2\font plus
\BIBentryALTinterwordstretchfactor\fontdimen3\font minus
  \fontdimen4\font\relax}
\providecommand{\BIBforeignlanguage}[2]{{%
\expandafter\ifx\csname l@#1\endcsname\relax
\typeout{** WARNING: IEEEtran.bst: No hyphenation pattern has been}%
\typeout{** loaded for the language `#1'. Using the pattern for}%
\typeout{** the default language instead.}%
\else
\language=\csname l@#1\endcsname
\fi
#2}}
\providecommand{\BIBdecl}{\relax}
\BIBdecl

\bibitem{Eilat_us}
M.~Yemini, A.~{Somekh-Baruch}, and A.~Leshem, ``On channels with asynchronous
  state information at the transmitter,'' \emph{Electrical and Electronics
  Engineers in Israel (IEEEI), 2012 IEEE 27th Convention of}, 2012.

\bibitem{shannon}
C.~E. Shannon, ``Channels with side information at the transmitter,'' \emph{IBM
  Research and Development}, vol.~2, pp. 289--293, 1958.

\bibitem{Kusnetsov}
A.~Kusnetsov and B.~Tsybakov, ``Coding in a memory with defective cells,''
  \emph{Problemy Peredachi Informatsii}, vol.~10, no.~2, pp. 52--60, 1974.

\bibitem{GP}
S.~I. Gelf'and and M.~S. Pinsker, ``Coding for channel with random
  parameters,'' \emph{Problem of Control and Information Theory}, vol.~9,
  no.~I, pp. 19--31, 1980.

\bibitem{Heegard}
C.~Heegard and A.~E. Gamal, ``On the capacity of computer memories with
  defects,'' \emph{IEEE Trans. Inf. Theory}, vol. IT-29, no.~5, pp. 731--739,
  September 1983.

\bibitem{Devroye}
N.~Devroye, P.~Mitran, and V.~Tarokh, ``Achievable rates in cognitive radio
  channels,'' \emph{IEEE Trans. Inf. Theory}, vol.~52, no.~5, pp. 1813--1827,
  May 2006.

\bibitem{GoldsmithJafar2009}
A.~Goldsmith, A.~Jafar, I.~Mari\'{c}, and S.~Srinivasa, ``Breaking spectrum
  gridlock with cognitive radio: An information theoretic perspective,''
  \emph{Proceedings of the IEEE}, vol.~97, no.~5, pp. 894--914, May 2009.

\bibitem{Moulin}
P.~Moulin and J.~A. O'Sullivan, ``Information-theoretic analysis of information
  hiding,'' \emph{IEEE Trans. Inf. Theory}, vol.~49, no.~3, pp. 563--593, March
  2003.

\bibitem{Caire}
G.~Caire and S.~{Shamai (Shitz)}, ``On the achievable throughput of a
  multiple-antenna {G}aussian broadcast channel,'' \emph{IEEE Trans. Inf.
  Theory}, vol.~49, no.~7, pp. 1691--1706, July 2003.

\bibitem{Anelia2008}
A.~{Somekh-Baruch}, S.~{Shamai (Shitz)}, and S.~Verd{\' u}, ``Cooperative
  multiple access encoding with states available at one transmitter,''
  \emph{IEEE Trans. Inf. Theory}, vol.~54, pp. 4448--4469, October 2008.

\bibitem{CoverMcEliece1981}
T.~M. Cover, R.~J. {McEliece}, and E.~C. Posner, ``Asynchronous multiple-access
  channel capacity,'' \emph{IEEE Trans. Inf. Theory}, vol.~27, no.~4, pp.
  409--413, July 1981.

\bibitem{HuiHumblet1985}
J.~Hui and P.~A. Humblet, ``The capacity region of the totally asynchronous
  multiple-access channel,'' \emph{IEEE Trans. Inf. Theory}, vol.~31, no.~2,
  pp. 207--216, March 1985.

\bibitem{Verdu1989}
S.~Verd\'{u}, ``The capacity region of the symbol-asynchronous {G}aussian
  multiple-access channel,'' \emph{IEEE Trans. Inf. Theory}, vol.~35, no.~4,
  pp. 733--751, July 1989.

\bibitem{Khisti}
A.~Khisti, U.~Erez, A.~Lapidoth, and G.~Wornell, ``Carbon copying onto dirty
  paper,'' \emph{IEEE Trans. Inf. Theory}, vol.~53, no.~5, pp. 1814--1827, May
  2007.

\bibitem{Piantanida}
P.~Piantanida and S.~Shamai{ (Shitz)}, ``Capacity of compound state-dependent
  channels with states known at the transmitter,'' in \emph{Proc. IEEE Int.
  Symp. Information Theory (ISIT'10)}, Seoul, Korea, 2010, pp. 624--628.

\bibitem{Nair}
C.~Nair, A.~{El-Gamal}, and Y.~K. Chia, ``An achievability scheme for the
  compound channel with state noncausally available at the encoder,'' \url{
  arXiv:1004.3427}, April 2010.

\bibitem{WeissmanElGamal}
T.~Weissman and A.~E. Gamal, ``Source coding with limited-look-ahead side
  information at the decoder,'' \emph{IEEE Trans. Inf. Theory}, vol.~52,
  no.~12, pp. 5218--5239, December 2006.

\bibitem{Wolfowitz1978}
J.~Wolfovitz, \emph{Coding Theorems of Information Theory}.\hskip 1em plus
  0.5em minus 0.4em\relax New York: Springer-Verlag, 1978.

\bibitem{HeegardElGamal1983}
C.~Heegard and A.~{El-Gamal}, ``On the capacity of computer memory with
  defects,'' \emph{IEEE Trans. Inf. Theory}, vol.~29, no.~5, pp. 731--739,
  September 1983.

\bibitem{Salehi1992}
M.~Salehi, ``Capacity and coding for memories with real-time noisy defect
  information at encoder and decoder,'' \emph{Proc, Inst. Elec. Eng.}, vol.
  132, no.~2, April.

\bibitem{GoldsmithVaraiya1997}
A.~J. Goldsmith and P.~P. Varaiya, ``Capacity of fading channels with channel
  side information,'' \emph{IEEE Trans. Inf. Theory}, vol.~43, no.~6, pp.
  1986--1992, November 1997.

\bibitem{Rosenzweig2005}
A.~Rosenzweig, Y.~Steinberg, and S.~{Shamai (Shitz)}, ``On channels with
  partial channel state information at the transmitter,'' \emph{IEEE Trans.
  Inf. Theory}, vol.~51, no.~5, pp. 1817--1830, May 2005.

\bibitem{Mitola}
J.~Mitola, ``Cognitive radio: An integrated agent architecture for software
  defined radio,'' Ph.D. dissertation, KTH Royal Institute of Technology
  Stockholm, Sweden, 2000.

\bibitem{Haykin}
S.~Haykin, ``Cognitive radio: Brain-empowered wireless communications,''
  \emph{IEEE Journal on Selected Areas in Communication}, vol.~23, no.~2, pp.
  201--220, February 2005.

\bibitem{Mitrpant2006}
C.~Mitrpant, A.~J.~H. Vinck, and Y.~Luo, ``An achievable region for the
  {G}aussian wiretap channel with side information,'' \emph{IEEE Trans. Inf.
  Theory}, vol.~52, no.~5, May.

\bibitem{ChenVinck2008}
Y.~Chen and A.~J.~H. Vinck, ``Wiretap channel with side information,''
  \emph{IEEE Trans. Inf. Theory}, vol.~54, no.~1, January.

\bibitem{SimeoneYener2009}
O.~Simeone and A.~Yener, ``The cognitive multiple access wire-tap channel,'' in
  \emph{Proc. 43rd Annu. CISS}, March 2009, pp. 158–--163.

\bibitem{KhistiDiggavi2011}
A.~Khisti, S.~N. Diggavi, and G.~W. Wornell, ``Secret-key agreement with
  channel state information at the transmitter,'' \emph{IEEE Trans. Inf.
  Forensics Security}, vol.~6, no.~3, September.

\bibitem{Xu2014}
P.~Xu, Z.~Ding, X.~Dai, and K.~K. Leung, ``A general framework of wiretap
  channel with helping interference and state information,'' \emph{IEEE Trans.
  Inf. Forensics Security}, vol.~9, no.~2, February.

\bibitem{Karakayali2006}
M.~K. Karakayali, G.~J. Foschini, and R.~A. Valenzuela, ``Network coordination
  for spectrally efficient communications in cellular systems,'' \emph{IEEE
  Wireless Communications}, vol.~13, pp. 56--61, August 2006.

\bibitem{Irmer2011}
R.~Irmer, H.~Droste, P.~Marsch, M.~Grieger, G.~Fettweis, S.~Brueck, H.~P.
  Mayer, L.~Thiele, and V.~Jungnickel, ``Coordinated multipoint: Concepts,
  performance, and field trial results,'' \emph{IEEE Communications Magazine},
  vol.~49, pp. 102--111, February 2011.

\bibitem{LiuLi2013}
Y.~Liu, Y.~Li, D.~Li, and H.~Zhang, ``Space-time coding for time and frequency
  asynchronous {CoMP} transmissions,'' \emph{IEEE Wireless Communications and
  Networking Conference (WCNC)}, pp. 2632--2637, April 2013.

\bibitem{us_ISIT_2014}
M.~Yemini, A.~{Somekh-Baruch}, and A.~Leshem, ``On the asynchronous cognitive
  {MAC},'' in \emph{Proc. IEEE Int. Symp. Information Theory (ISIT'14)},
  Honolulu, HI, USA, June/July 2014, pp. 2929--2933.

\bibitem{us_paper2}
------, ``On the multiple access channel with asynchronous cognition,''
  \emph{IEEE Trans. Inf. Theory - in preparation}, 2014.

\bibitem{CT}
T.~M. Cover and J.~A. Thomas, \emph{Elements of information theory},
  2nd~ed.\hskip 1em plus 0.5em minus 0.4em\relax Wiley Interscience, 2006.

\bibitem{Merhav}
N.~Merhav and T.~Weissman, ``Coding for the feedback {G}el'fand-{P}insker
  channel and the feedforward {W}yner-{Z}iv source,'' \emph{IEEE Trans. Inf.
  Theory}, vol.~52, no.~9, pp. 4207--4211, September 2006.

\bibitem{maric}
I.~Mari{\'c}, N.~Liu, and A.~Goldsmith, ``Encoding against an interferer's
  codebook,'' in \emph{Proc. Allerton Conf. Communications, Control, and
  Computing}, Monticello, IL, September 2008, pp. 523--530.

\bibitem{NetworkInformationTheory}
A.~{El-Gamal} and Y.~H. Kim, \emph{Network Information Theory}.\hskip 1em plus
  0.5em minus 0.4em\relax Cambridge University Press, 2011.

\bibitem{Kramer}
G.~Kramer, ``Topics in multi-user information theory,'' \emph{Foundations and
  Trends in Communications and Information Theory}, vol.~4, no. 4--5, pp.
  265--444, 2007.

\end{thebibliography}
\end{document}